\documentstyle[11pt]{article}
\makeatletter

\if@twoside
\else
\fi
\ifcase \@ptsize
\or 
\or 
\fi

\@addtoreset{equation}{section}
\def\section{\@startsection {section}{1}{\z@}{-3.5ex plus -1ex minus
 -.2ex}{2.3ex plus .2ex}{\large\bf\centering}}
\def\subsection{\@startsection{subsection}{2}{\z@}{-3.25ex plus%
 -1ex minus -.2ex}{1.5ex plus .2ex}{\bf}}
\def\subsubsection{\@startsection{subsubsection}{3}{\z@}{-3.25ex plus%
 -1ex minus -.2ex}{1.5ex plus .2ex}{\sl}}
\gdef\@publabel{\hfil}
\gdef\@pubdate{\null}
\gdef\@pubnumber{\null}
\gdef\@author{\null}
\gdef\@title{\null}
\gdef\@abstract{\null}
\long\def\pubdate#1{\gdef\@pubdate{#1}}
\long\def\pubnumber#1{\gdef\@pubnumber{#1}}
\long\def\publabel#1{\gdef\@publabel{#1}}
\long\def\author#1{\gdef\@author{#1}}
\long\def\title#1{\gdef\@title{#1}}
\long\def\abstract#1{\gdef\@abstract{#1}}
\def\titlerelax{
\let\maketitle\relax
\let\settitleparameters\relax
\let\consolidatetitle\relax
\let\inittitlepage\relax
\let\finishtitlepage\relax
\let\titlepagecontents\relax
\let\multithanks\relax
\let\titlebaselines\relax
\let\@makepub\relax
\let\@maketitle\relax
\let\@makeauthor\relax
\let\@makeabstract\relax
\let\@maketitlenote\relax
\let\thanks\relax
\let\titlerelax\relax}
\def\titleclean
{\gdef\@titlenote{}
\gdef\@abstract{}
\gdef\@author{}
\gdef\@title{}
\gdef\@pubdate{}\gdef\@pubnumber{}\gdef\@publabel{}
\gdef\@dpublabel{}
}

\def\@makepub{\vbox to \z@{\hbox to \textwidth{\hfill
\@publabel \hfill
\llap{\parbox[t]{0.33\textwidth}{\raggedleft\@pubnumber}}}%
\vss}}
\def\@maketitle{\vskip 60pt \begin{center}
 {\LARGE \@title \par}
 \end{center}}

\def\@makeauthor{{%
\def\and{\smallskip {\normalsize \rm and\smallskip }}
\def\And{\medskip {\normalsize \rm and\\}\medskip}
\long\def\address##1{{\def\and{\\and\\}\medskip
                                {\small \it \\##1\\}
}}
{\centering
 \vskip 3em
 \large \lineskip .75em
 \@author}
 \par}}
\def\@makedate{\vskip 1.5em
 {\raggedright \small \noindent\@pubdate \par}}
\def\@makeabstract{\vskip 1.5em
{\small
\begin{center}
{\bf ABSTRACT\vspace{-.5em}\vspace{0pt}}
\end{center}
\quotation \@abstract \endquotation}}
\def\maketitle{\titlepage
\let\footnotesize\small \setcounter{page}{0}
\@makepub
\vfil
\@maketitle
\@makeauthor
\vfil
\@makeabstract
\@thanks
\vfil
\@makedate
\if@restonecol\twocolumn \else \eject \fi
\titlerelax \titleclean
\setcounter{footnote}{0}
}

\newcommand{\ncm}{\newcommand}

\ncm{\be}{\begin{equation}
\addtolength{\abovedisplayskip}{\extraspaces}
\addtolength{\belowdisplayskip}{\extraspaces}
\addtolength{\abovedisplayshortskip}{\extraspace}
\addtolength{\belowdisplayshortskip}{\extraspace}}
\ncm{\ee}{\end{equation}}

\ncm{\bea}{\begin{eqnarray}
\addtolength{\abovedisplayskip}{\extraspaces}
\addtolength{\belowdisplayskip}{\extraspaces}
\addtolength{\abovedisplayshortskip}{\extraspace}
\addtolength{\belowdisplayshortskip}{\extraspace}}
\ncm{\eea}{\end{eqnarray}}

\ncm{\beas}{\begin{eqnarray*}
\addtolength{\abovedisplayskip}{\extraspaces}
\addtolength{\belowdisplayskip}{\extraspaces}
\addtolength{\abovedisplayshortskip}{\extraspace}
\addtolength{\belowdisplayshortskip}{\extraspace}}
\ncm{\eeas}{\end{eqnarray*}}

\newlength{\extraspace}
\newlength{\extraspaces}
\def\blp{{\bar\lambda + \bar\rho}}

\def\bm{{\bar\mu}}

\def\cev#1{{ \langle {#1} \vert }}
\def\cF{{\cal F}}
\def\cH{{\cal H}}
\def\ch{{\rm ch}}

\def\hs#1{{\hat s_{#1}}}

\def\lpp{{\lambda + \rho}}
\def\PHI#1#2#3#4{ \Phi({\ss{{#1\,#2}\atop{#3\,#4}}}) }
\def\PHI#1#2#3#4{ \Phi({\tiny\begin{array}{cc}
{\!\!#1\!}&{\!#2\!\!}\\
{\!\!#3\!}&{\!#4\!\!}\end{array}}) }
\def\PHI#1#2#3#4{{\phi_{[#1\,#2;#3\,#4]}}}

\def\rn #1{{\romannumeral #1}}
\def\rn #1{\hbox{\rm\expandafter\uppercase\expandafter{\romannumeral #1}}}

\def\slt{{a_2^{(1)}}}

\def\vac{{\vec 0}}

\def\vec#1{{ \vert {#1} \rangle }}

\def\WA{{W_3}}     
\def\WB{{W(2,4)}}  

\begin{document}

\pubnumber{LAVAL-PHY-96-20 \\ CRM-2313 \\ DAMTP--96--20}
\pubdate{March 13, 1996}

\title{
Probing integrable perturbations of conformal theories using singular
vectors
}

\author{Pierre Mathieu
\address{D\'epartement de
Physique, Universit\'e Laval, Qu\'ebec, Canada G1K 7P4}%
\medskip\And\medskip
Gerard Watts\thanks{Address from March 1, 1996: Department of
Mathematics, King's College London, The Strand, London, WC2R 2LS}
\address{Centre de Recherches Math\'ematiques,
Universit\'e de Montr\'eal, Qu\'ebec, Canada H3C 3J7
\and
D.A.M.T.P., University of Cambridge, Silver Street, Cambridge
CB3 9EW, U.K.}
}

\abstract{
It has been known for some time that the $(1,3)$ perturbations of the
$(2k+1,2)$ Virasoro minimal models have conserved currents which are
also singular vectors of the Virasoro algebra. This also turns out
to hold for  the $(1,2)$ perturbation of the $(3k\pm1,3)$ models. In this
paper we investigate the requirement that a perturbation of an
extended conformal field theory has conserved currents which are also
singular vectors. We consider conformal field theories with $\WA$ and
(bosonic) $WBC_2=\WB$  extended symmetries.
Our analysis relies heavily on the general conjecture of de Vos and
van Driel relating the multiplicities of $W$-algebra irreducible
modules to the Kazhdan-Lusztig polynomials of a certain double
coset. Granting this conjecture, the singular-vector argument provides
a direct way of recovering all known integrable
perturbations. However, $W$  models bring  a slight complication in
that the conserved densities of some $(1,2)$-type perturbations are
actually subsingular vectors, that is, they become singular vectors
only in a quotient module.
}

\maketitle

\section{Introduction}

When perturbed by a relevant operator $\Phi(z,{\bar z})$, 
a conformal field theory will no longer remain conformally
invariant and there will be massive particles in the spectrum.
In some instances, the complete
integrable structure present at the critical point is preserved by the
perturbation \cite{ZAMO}, i.e.,\, an infinite number of mutually
commuting integrals remain conserved off-criticality in the massive
theory. An integral of the form $H_n= \int {\cal H}_{n+1} dz$, where
${\cal H}_{n+1}$ is a function of the holomorphic conserved currents
of the conformal field theory, is said to remain conserved in the
massive regime if $\partial_{{\bar z}} H_n=0$. To first order in
perturbation theory, to which our analysis is confined, this is
equivalent to
\be
  [H_n, \int\!\Phi~dz] = 0
\;.
\label{eq:crit}
\ee
This condition is satisfied when the residue of the operator
product  ${\cal H}_{n+1}(z)\Phi(w,{\bar w})$ is a total derivative.

Feigin and Frenkel \cite{FF} have shown that the perturbations of a
conformal field theory by Virasoro primary fields of types 
$\{ \Phi_{1,2}, \Phi_{1,3}, \Phi_{1,5}, 
    \Phi_{2,1}, \Phi_{3,1}, \Phi_{5,1} \}$
are integrable by criterion (\ref{eq:crit}) for generic values of $c$
the central charge. Furthermore, these perturbations may be grouped
into dual pairs 
$\{ \Phi_{1,2}, \Phi_{5,1} \}$, 
$\{ \Phi_{1,5}, \Phi_{2,1} \}$, 
$\{ \Phi_{1,3}, \Phi_{3,1} \}$, 
for which the pairs have the same conserved densities.
For any particular value of $c$ only some subset of these 6
perturbations will be  relevant, i.e., the perturbing field has
dimension less than 2.
For the unitary minimal models, the relevant 
integrable  perturbations are $\Phi_{1,3}$, $\Phi_{1,2}$ and
$\Phi_{2,1}$. The conservation laws of the corresponding perturbed
theory are exactly those of the quantum version of the Korteweg-de
Vries (KdV) equation, the Gibbon-Saweda-Kotera (GSK) equation and the
Kuperschmidt (K) equation respectively \cite{KM}. The last two
equations are reductions of the quantum Boussinesq equation (for the
classical derivation, see \cite{FG}). The Toda relatives of these
three integrable systems are the sine-Gordon equation, based on the
$a_1^{(1)}$ affine algebra, and the  Bullough-Dodd equation, based on
the twisted $a_2^{(2)}$ affine algebra \cite{HM}. The existence of two
integrable reductions of the Boussinesq equation is rooted in the
asymmetry of the  $a_2^{(2)}$ Dynkin diagram.

For some nonunitary models, the perturbation $\Phi_{1,5}$ is also
relevant \cite{RSTa}. 
 As remarked on above, 
the conservation laws of the resulting massive theory 
 would also commute with a perturbation by $\Phi_{2,1}$, but 
the ranges of $p$ and $q$ for which they
are relevant are complementary: for the $(p,q)$ models, $\Phi_{1,5}$
 and $\Phi_{2,1}$ are relevant for $2q<p$ and $p<2q$
respectively.

This list of integrable perturbations is 
 believed to be complete.
The evidence supporting this assertion is the following (see
e.g.,\cite{KM,Ma}). There are exactly three distinct integrable
hierarchies (KdV, GSK and K) which are Hamiltonian with Poisson
brackets given by the classical version of the conformal algebra. The
integrability of their quantum versions readily implies that, at the
critical point, there are three distinct towers of commuting
integrals, whose densities are differential polynomials in the
energy-momentum tensor.  (The conserved integrals of each tower
commute among themselves but those in different towers do not
commute.)  Each sequence commutes with one and only one of the three
dual pairs of perturbations $\{ \Phi_{1,3}, \Phi_{3,1} \}$, $\{
\Phi_{1,2}, \Phi_{5,1} \}$ and $\{ \Phi_{1,5}, \Phi_{2,1} \}$, so that
each of these perturbations preserves an infinite number of conserved
integrals. That there are no further classical integrable hierarchies
which are Hamiltonian with respect to  the  `KdV second Hamiltonian
structure' is a strong argument for the absence of further towers of
quantum conserved integrals, hence of additional integrable
perturbations. 

In the context of the Yang-Lee model, for which the integrable
perturbation is $\Phi_{1,3}=\Phi_{1,2}$, it was observed in
\cite{FREUND} that some conserved densities become exactly
proportional to singular  vectors.  This observation has been proved
to be generic to all $(p,q)=(2k+1,2)$ models perturbed by the
$\Phi_{1,3}$ field \cite{DiMa,EY,KNS}: in these models, the
density $\cH_{2k}$ is proportional to the singular vector of dimension
$2k$ in the vacuum module. This singular vector is $\Phi_{3,1}$.  For
the corresponding value of the central charge, the conservation law of
dimension $2k-1$ is thus absent.  But, as argued in  \cite{DiMa}, the
important point remains that before modding out the  theory by the
singular vector, this latter provides a nontrivial conserved quantity
of the off-critical theory. Changing the value of $k$, hence the value
of the central charge $c$, produces singular conserved densities of
different dimensions.  This singular vector argument provides thus a
simple and powerful signal of the  integrability of the perturbed
theory. Notice moreover that the exact expressions for these conserved
densities are known since there are closed expressions for all the
$(m,1)$ type singular vectors of the Virasoro algebra  \cite{BSA}.

This argument generalises to the $(4k,2)$  superconformal minimal
models perturbed by the supersymmetric analogue of the $\Phi_{1,3}$
field and for a particular perturbation of the \hbox{$N=2$}
superconformal unitary models \cite{DiMa}. It has also been checked
explicitly  for the $\Phi_{[11;22]}$ perturbation of the $W_3$ minimal
model.  The same conclusion can be expected for all
$W_n\equiv WA_{n-1}^{(1)}$ minimal models in the sequence
$(p,q)=(p,n)$ (with $q$ coprime to $n$)  thanks to the KNS duality:
$W_n^{(p,n)}= W_{p-n}^{(p,p-n)}$ \cite{KNS}.

In section \ref{sec:two}, we reconsider the perturbations of Virasoro
minimal models. We consider the possibility of probing integrable
perturbations other than  $\Phi_{1,3}$ by explicitly requiring the
perturbation to have a conserved density which is also a singular
vector of the Virasoro algebra. We recover exactly the integrable
perturbations already known. A simple free-field argument is presented
which leads to this restriction, but a more detailed analysis is
needed to show the existence of the singular conserved densities.

Next we analyse the perturbations of $W$-minimal models.
We first recall in appendix \ref{sec:six} the connection with free-field
constructions and Toda theories, and find that the simple idea used in
section \ref{sec:two} to explain the short list of integrable
perturbations also only works for the perturbations related to affine
Toda theories. However, the free-field argument only indicates when
there may be singular conserved quantities; we have to perform a more
sophisticated analysis to show that there really are singular
conserved quantities. This analysis is presented in sections
\ref{sec:wa} and \ref{sec:wb} for the $\WA$ and $\WB$ algebras
respectively.

The free-field analysis again yields dual pairs of fields; 
given a dual pair, we can regard any of the two fields as the
perturbing field, the other one being the conserved density.  
In the Virasoro case, up to a duality transformation, there is always
a value of $c$ such that this conserved density is in a  one--to--one
correspondence with a genuine singular vector   in the vacuum module.
However, the complicated structure of  $W$ modules allows for another
possibility: the conserved density can be associated to a  subsingular
vector, namely a vector that becomes singular only in a quotient
module. This is actually realised for some nonminimal models with a
rational central charge.  The relevant quotient module, denoted by
$\cF $, is the vacuum module quotiented by the submodule generated by
the states $L_{-1}\vac$ and $W_{-1}\vac$ -- this is equivalent to the
space of fields which are normal ordered combinations of $L$, $W$ and
their derivatives. This complication forces us into a detailed
analysis of the quotient space $\cF $. This analysis is presented in
sections \ref{sec:kl} and \ref{sec:wba} for the $\WA$ and $\WB$ cases
respectively. In section \ref{sec:search}, we show that certain $\WA$
perturbations  are  integrable for particular values of $c$ by
demonstrating that their conserved densities are singular vectors in
$\cF $; a similar analysis is presented in section \ref{sec:seven} for
the $\WB$ case. Once the integrability of a given perturbation is
established, by relating the conserved density to a vacuum
(sub)singular vector, the integrability of the perturbing field that
can be obtained by duality follows automatically.  

Explicit computations of conserved charges for the perturbation
under consideration are reported in section \ref{sec:nine} and
compared with those predicted by the singular-vector analysis.

We would like to stress that for all the perturbations we find to
be integrable at a discrete set of values of the central charge, the
existence of an infinite set of conserved charges for all values of
$c$ has been proven  in  \cite{FF}. Each perturbed conformal field
theory we consider can be thought of as an affine Toda theory, and in
\cite{FF} it is  proven that, in the quantum affine Toda theories,
there are conserved charges with spins equal to the exponents of the
corresponding affine algebras, although their proof is not
constructive.

The main result of this paper is that for the Virasoro, $\WA$ and
$\WB$ algebras, there are no new integrable perturbations for which the
conserved densities are singular vectors, and that all known
integrable perturbations may be found from this criterion.

\section{Perturbed Virasoro minimal models}
\label{sec:two}

\subsection{The singular-vector argument}

By Virasoro minimal model, we mean a conformal field theory of central
charge $c(p,p')$
\be
  c(p,p') = 1 - 6 \frac{(p-p')^2}{pp'}
\ee
containing the set of fields $\Phi_{m,n; \bar m,\bar n}(z,\bar z)$
which transform under the holomorphic (resp. anti-holomorphic)
Virasoro algebra as primary fields of conformal weight $h_{m,n}$
(resp. $h_{\bar m, \bar n}$):
\be
  h_{m,n}
= {(mp-np')^2-(p-p')^2\over 4pp'}
\;
\label{eq:vir}
\ee
and where $m,n$  (and $\bar m,\bar n$) satisfy
\be
  1\leq m \leq p'-1 \,,\; 1\leq n \leq p-1
\;.\;\;
\label{eq:mnrange}
\ee
The values of $h_{m,n}$ satisfying (\ref{eq:mnrange}) are said to form
the Kac table of the $(p,p')$ minimal model%
\footnote{
Although the field content (i.e.\ the sets $\{m,n;\bar m,\bar n\}$
actually occurring) of a Virasoro minimal model can be fixed by
enforcing the modular invariance of the  partition function for the
theory defined on a torus,  this does not seem the relevant
requirement for massive integrable models defined on a line
(see e.g.~\cite{Anni} for some interesting developments in this area.)
}.

We consider a perturbed conformal field theory with perturbing field%
\footnote{
Observe that the antiholomorphic labels need not be the same as the
holomorphic ones.  When  $h_{m,n}\not=h_{\bar m, \bar n}$, the
perturbation is not spinless and it breaks Lorentz invariance.  For
instance the chiral Potts model is a perturbation of the Potts model
by the integral of a spin one field \cite{Potts}
}
\be
     \Phi_{m,n; \bar m,\bar n}
\sim \phi_{m,n}(z){\bar \phi}_{\bar m,\bar n}({\bar z})
\;,
\ee
In this perturbed theory, we look for a conserved quantity
whose density is also a singular vector. Since the conserved
density is a polynomial in the energy-momentum tensor $L(z)$ and its
derivatives, it is independent of $\bar z$. Moreover, since it is a
singular vector, it is also a Virasoro primary field.  It is thus
necessarily of the form
\be
\Phi_{r,s;1,1}(z,\bar z) \equiv \phi_{r,s}(z)
\;,
\ee
where now $\{r,s\}$ lie outside the Kac table (since it is also a descendant
field).

Usually,  all such singular vector descendants $\phi_{m',n'}$ 
 of primary fields $\phi_{m,n}$ are 
identically set equal to zero. However, we may instead consider a
partially reduced set of fields, in which only the level  $mn$
singular vector  descendant in  the representation $(m,n)$ is set
equal to zero%
\footnote{Since in Liouville theory this singular vector at level $mn$
is proportional to the equations of motion for the perturbing field,
it should indeed be set to zero identically.
}
(e.g. for the vacuum module, this means that  $L_{-1}\vac \equiv 0$
identically) but any null state that is not a descendant of this
level $mn$ singular vector is not set to zero (yet).

The structure of Virasoro algebra representations is very well
understood \cite{FeFu2}: if a singular vector in the vacuum module is
not a descendant of $L_{-1}\vac$, then it can only correspond to the
non-trivial leading singular vector in the vacuum Verma module of a
minimal model. The corresponding field  is of type
$\phi_{2p'-1,1} \equiv \phi_{1,2p-1}$
and it transforms as a primary field of weight $(p-1)(p'-1)$.
Similarly, the only singular vector $\phi_{m',n'}$ which is a
descendant of $\phi_{m,n}$, but not of the singular vector at level
$(mn)$, must be at level $(p'-m)(p-n)$ and transform as a primary field
of type $\phi_{2p'-m,n} \equiv \phi_{m,2p-n}$.

In order to decide whether this field is indeed a conserved density
for the perturbed theory, we need to investigate the ope
$\phi_{r,s}(z)  \phi_{m,n}(w)$.
The fusion rules obtained in \cite{BPZ} are only applicable to fields
lying in the Kac table and thus, they cannot  be applied directly to
our case.
In \cite{FeFu3}, Feigin and Fuchs have determined those pairs of
fields that can fuse with a given  third field by requiring the
decoupling of  all null-vector descendants of the third field in
three-point functions with the first two fields. Their argument is
rather involved, however, and in our simple cases we can use the
following alternative argument.
First, since $\phi_{r,s}(z)$ is a polynomial in $L(z)$ 
the only primary fields $\phi_{m',n'}$ which may occur in the o.p.e. 
\be
\phi_{r,s}(z) \;\; \phi_{m,n}(w) \sim \sum_{m',n'}\phi_{m',n'}
\;,
\ee
are Virasoro-descendants of $\phi_{m,n}$. $\phi_{m,n}$ itself cannot
occur, since 
the singular vector $\phi_{r,s}$  decouples from the physical states in the
theory.

Consequently, the only o.p.e. we have to consider 
 in our theory is
\be
  \phi_{2p'-1,1}(z) \;\;
  \phi_{m,n}(w) \;
\sim (z-w)^{-\Delta h} \phi_{2p'-m,n}(w) \; + \; \ldots
\;,
\label{eq:ope1}
\ee
where $\{m,n\}$ are in the Kac table (\ref{eq:mnrange}) --
since these are the labels of the perturbing field -- and
\be
  \Delta h
= h_{2p'-1,1} + h_{m,n} - h_{2p'-m,n}
= (p-n)(m-1) + (p'-m)(n-1) + (m-1)(n-1) \geq 0
\;.
\ee
We may determine the whole o.p.e. (\ref{eq:ope1}) by solving the descent
equations in \cite{BPZ},
\[
  \phi_{h_1}(z)   \vec{h_2}  \;
\propto
   z^{h_3 - h_1 - h_2} \Big[   
  1 
+ z \frac{ h_1 + h_3 - h_2 }{ 2 h_3 } L_{-1}
+ z^2 \big(
\frac{ (h_1 + h_3 - h_2 )(h_1 + h_3 - h_2 + 1 ) }
     { 4 h_3 ( 2 h_3 + 1 )}       L_{-1}^2 
\]
\be
+ \frac{ (h_1 + h_2)(2 h_3 + 1)+ h_3( h_3 - 1) -  3( h_1 - h_2)^2 }
       { 16 h_3^2 + (2c - 10) h_3 + c}
  \left[ L_{-2} - \frac{3}{2(h_3 + 1)} L_{-1}^2 \right]
  \big)
+ \ldots  \Big]\vec{h_3}
\;.
\label{eq:ope2} 
\ee
Clearly the residue of the pole in (\ref{eq:ope2}) will be a total
derivative if $\Delta h =h_1+h_2-h_3= 2$. We leave the possibility that 
$\Delta h >2 $ to the end of this section.

Requiring $\Delta h =2$ and $\{m,n\}$ in the Kac table gives only two
solutions  (taking $p \geq p'$)
\be
  \begin{array}{l@{\;\;}l}
   (a) & m=1\;, n=2\;, p'=3 \;,   \\
   (b) & m=1\;, n=3\;, p'=2 \;.
  \end{array}
\ee
In case (a) the conserved current is $\phi_{5,1}$. Since $p'=3$,
$p=3k\pm 1$ and the conservation laws have dimension $h_{5,1}-1=6k-1,
6k-5$  (as in the quantum GSK case)%
\footnote{The relation between the conserved densities of the $(1,2)$
perturbed theory and the $\phi_{(5,1)}$ singular vectors was first
noticed by P. Di Francesco and F. Lesage (unpublished).
}.

In case (b) the conserved current is $\phi_{3,1}$. Since $p'=2$,
$p=2k+1$, $h_{3,1}=2k$ and the conservation laws have dimension $2k-1$
(as in the  quantum KdV case).

Note that, by a  duality argument, the integrability of the $(1,2)$
perturbation implies that of $(1,5)$%
\footnote{
The singular-vector argument does not apply directly to the $(1,5)$
perturbation since $(2,1)$ can never be in the vacuum module.
}.
This argument, evident in the free field construction but rather
mysterious otherwise, runs as follows:

In the free field approach, a primary field $\phi_{m,n}$ can be
represented by
\be
  \exp( - i \frac{ (m-1)p - (n-1)p'}{\sqrt{ 2 pp'}} \varphi )
\;,
\label{eq:ff}
\ee
where $\varphi$ is a free  boson. If a quantity of the form
$\int \cH ~dz$ (where the field $\cH$ is polynomial in the field $L(z)$
and its derivatives) commutes with
$\int\exp(i \gamma \varphi)~dz$, then it also commutes with the operator
$\int\exp(-i 2 \varphi / \gamma )~dz$
\cite{FF,KWat3}.
Hence, if $\int\cH ~dz$ commutes with $\int \phi_{1,m}~dz$, it also commutes
$\int \phi_{ (m+3)/(m-1) , 1}$. The only positive integers $m$ for which
$(m+3)/(m-1)$ is also an integer are 2, 3 and 5.
As a result, since $\phi_{1,2}$ is an integrable perturbation, so is
$\phi_{5,1}$, and by interchange of $p$ and $p'$, so is $\phi_{1,5}$.

Hence, supplementing the singular-vector analysis with this
duality, we end up with the  following list of integrable
perturbations of the $\phi_{1,n}$ type:  $\phi_{1,3}, \phi_{1,2}$ and
$\phi_{1,5}$. Quite remarkably, our simple singular-vector argument
allows us to recover the complete list of integrable perturbations.
Thus, in all cases,  explicit expressions of the conservation laws are
known at some values of $c$.

\subsubsection{$\Delta h > 2$}

To complete our argument, we show that $\Delta h>2$ cannot lead to
further generic  solutions. Let us first consider $\Delta h = 3$.
Asking that $m,n,p,p'$ are positive integers with $p \geq p'$ we find
the possibilities 
\[
\begin{array}{c|cccc}
     & m & n & p & p' \\
\hline
(a)  & 1 & 2 &   & 4   \\ 
(b)  & 1 & 4 &   & 2   \\ 
(c)  & 2 & 4 & 4 & 2   \\ 
(d)  & 2 & 2 & 4 & 2   \\ 
(e)  & 2 & 3 & 4 & 2   \\ 
(f)  & 2 & 2 & 3 & 3   \\ 

\end{array}
\]
Since we require $p$ and $p'$ coprime, of these only $(a)$ and $(b)$
can correspond to sensible models. If we now ask that the residue of
the o.p.e. (\ref{eq:ope2}) is a total derivative with $\Delta h = 3$,
then the term  
\be
 \frac{ (h_1 + h_2)(2 h_3 + 1)+ h_3( h_3 - 1) -  3( h_1 - h_2)^2 }
       { 16 h_3^2 + (2c - 10) h_3 + c}
  \left[ L_{-2} - \frac{3}{2(h_3 + 1)} L_{-1}^2 \right]
  \vec{h_3}
\;, 
\label{eq:van} 
\ee
must vanish. This can happen, either because the coefficient vanishes
identically, or because the state is a singular vector (which is
indicated by the coefficient being singular).
The coefficient becomes 
\[
\begin{array}{cc}
(a)  & 
{\displaystyle 
\frac { 36p^4 - 132p^3 + 181p^2 - 132p + 36}
      {12 p (p - 2)(5p - 4)(3p - 4)}
}
\\
(b)  & 
{\displaystyle 
\frac {3( 4 p^4 - 36 p^3 + 129 p^2 - 270 p + 225 )}
      { 40(\,p - 3 )(p - 6)(p - 2)^2}
}
\\

\end{array}
\]
So we find that, again requiring $p$ and $p'$ coprime, the residue is
a total derivative with $\Delta h=3$ for the single case
\[
  m=1,n=4,p=3,p'=2
\;.
\]
In this case, however, $\phi_{m,2p-n}$ is not a descendant of
$\phi_{m,n}$, but the reverse is in fact true. This means that there
are no perturbations which have singular conserved densities with
$\Delta h = 3$.

Although we cannot prove that there are no sporadic cases where the
residue of the o.p.e. is a total derivative for $\Delta h>3$, we
certainly expect that these will not yield any new series of
integrable perturbations. 

\subsection{Explicit  calculations of conserved charges}

To verify our conclusions, we calculated $\cH_n$ for the $(1,2)$,
$(1,3)$ and $(1,5)$ perturbations for small $n$ and checked when these
fields were also primary fields. Being in the vacuum module, these
primary fields correspond then  necessarily to singular vectors. The
only values of $c$ for which this occurs are exactly those
predicted by the previous analysis.

The following table shows when a conserved quantity is found, and for which
values of the parameters the conserved current is a highest-weight
vector. $t$ stands for the ratio $p'/p$.
An `X' indicates that there is no conserved quantity.
A blank entry indicates that we did not consider this case.

In each case, the perturbation can be associated with an
affine Toda theory, as explained in appendix \ref{app:toda}.

{
\renewcommand{\arraystretch}{1.5}
\[
\begin{array}{||c||c|c|c||}
\hline \hline
n
& 
\begin{array}{c}
a_2^{(2)} \\
h_{1,2} = \frac{3}{4}t - \frac 12
\end{array}
&
\begin{array}{c}
a_1^{(1)} \\
h_{1,3} = 2 t - 1
\end{array}
&
\begin{array}{c}
{a_2^{(2)}} \\
h_{1,5} = 6 t - 2
\end{array}
\\[1mm] \hline\hline
  3
&
  {\rm X}
&
  \left\{
  {\renewcommand{\arraystretch}{1.1}
  \begin{array}{l} 
  t = 5/2 \\ c = -22/5 \\ h  = 4 
  \end{array}}
  \right.
  \left\{
  {\renewcommand{\arraystretch}{1.1}
  \begin{array}{l} 
  t = 2/5 \\ c = -22/5 \\  h = -1/5
  \end{array}}
   \right.
&
  {\rm X}
\\ \hline
  5
&
  \left.
  {\renewcommand{\arraystretch}{1.1}
  \begin{array}{l} 
  t = 3/4 \\ c = 1/2  \\ h = 1/16 \\
  \end{array}}
  \right.
&
  \left\{
  {\renewcommand{\arraystretch}{1.1}
  \begin{array}{l} 
  t = 7/2 \\ c = -68/7 \\ h= 6
  \end{array}} \right.
  \left\{
  {\renewcommand{\arraystretch}{1.1}
  \begin{array}{l} 
  t = 2/7 \\ c = -68/7 \\  h = -3/7
  \end{array}} \right.
&
  \left.
  {\renewcommand{\arraystretch}{1.1}
  \begin{array}{l} 
  t = 4/3 \\ c = 1/2 \\ h = 6 
  \end{array}} \right.
\\ \hline
  7
&
  \left.
  {\renewcommand{\arraystretch}{1.1}
  \begin{array}{l} 
  t = 3/5 \\ c = -3/5 \\  h = 1/5
  \end{array}} \right.
&
  \left\{
   {\renewcommand{\arraystretch}{1.1}
  \begin{array}{l} 
  t = 9/2 \\ c = -46/3 \\  h = 8 
  \end{array}} \right.
  \left\{
  {\renewcommand{\arraystretch}{1.1}
  \begin{array}{l} 
  t = 2/9 \\ c = -46/3 \\  h = -5/9 \\
  \end{array}}\right.
&
  \left.
   {\renewcommand{\arraystretch}{1.1}
  \begin{array}{l} 
  t = 5/3 \\ c = -3/5 \\  h = 8 \\ 
  \end{array}} \right.
\\ \hline
  9
&
  {\rm X}
&
  \left\{
  {\renewcommand{\arraystretch}{1.1}
  \begin{array}{l} 
  t = 11/2 \\ c = -232/11 \\ h = 10 \\
  \end{array}} \right.

  \left\{
  {\renewcommand{\arraystretch}{1.1}
  \begin{array}{l} 
  t = 2/11 \\ c = -232/11 \\ h = -7/11 \\
  \end{array}} \right.

&
  {\rm X}
\\ \hline
  11
&
  \left.
   {\renewcommand{\arraystretch}{1.1}
  \begin{array}{l} 
  t = 3/7 \\ c = -25/7 \\ h = -1/7 \\ 
  \end{array}} \right.

&
  \left\{
   {\renewcommand{\arraystretch}{1.1}
  \begin{array}{l} 
  t = 13/2 \\ c = -350/13 \\  h = 12 
  \end{array}} \right.

  \left\{
  {\renewcommand{\arraystretch}{1.1}
  \begin{array}{l} 
  t = 2/13 \\ c = -350/13 \\ h = -9/13 
  \end{array}} \right.

&
  \left.
  {\renewcommand{\arraystretch}{1.1}
  \begin{array}{l} 
  t = 7/3 \\ c = -25/7 \\ h = 12 \\
  \end{array}} \right.
\\ \hline
\end{array}
\]
}

\section{Integrable perturbations of the $\WA$ algebra }
\label{sec:wa}

\subsection{Introduction}

We first start with a brief review of the $\WA$ algebra. The
generators of the $\WA$ algebra  are denoted by $L_m$ and $ W_m$. They
satisfy the usual Virasoro relations and
\[
  [L_m,W_n]
= (2m-n)W_{m+n} \;,
\]
\[
  [W_m,W_n]
= {{(22 + 5c)}\over{48}} {{c}\over{3\cdot 5!}}
  (m^2-4)(m^2-1)m\delta_{m+n}
 + {1 \over 3}(m-n)\Lambda_{m+n}
\]
\[
 + \frac{(22 + 5c)}{48}\frac{(m-n)}{30}(2m^2-mn+2n^2-8)L_{m+n}
\;,
\]
where 
\[
  \vec\Lambda 
= (L_{-2}L_{-2} - (3/5)L_{-4})\vac
\;.
\]
This is related to the standard normalisation by
\[
  W = \sqrt{ ( 22 + 5c ) / 48 }\; W_{standard}
\;.
\]
The highest-weight representations are labelled by the central charge
$c$, and by $\{h,w\}$, the eigenvalues of $L_0$ and $W_0$.

The $W_3$ algebra is naturally derived as the quantum Hamiltonian
reduction of $a_2^{(1)}$ \cite{FKWa1,BTji}. The structure of the
highest-weight  representations is analysed in great detail in 
\cite{vddv} from this point of view.  Each highest-weight
representation of the $\WA$ algebra may be found as the quantum
Hamiltonian reduction of an associated $a_2^{(1)}$ highest-weight
representation with highest weight $\lambda$. To fix our notation, we
denote by $\{\alpha_0,\alpha_1,\alpha_2\}$ the $a_2^{(1)}$ simple
roots, with inner products $(\alpha_i,\alpha_j)=3\delta_{ij}-1$; the
set of positive real roots $\Delta^{re}_+$ of $a_2^{(1)}$ is 
\[
  \Delta^{re}_+ =
\{
\alpha_1 + n \delta
\;,\;\;
\alpha_1 + \alpha_2 + n \delta
\;,\;\;
\alpha_2 + n \delta
\;,\;\;
-\alpha_1 + (n+1) \delta
\;,\;\;
-\alpha_1 - \alpha_2 + (n+1) \delta
\;,\;\;
-\alpha_2 + (n+1) \delta
\}
\]
where $n=0,1,\ldots$ and $\delta = \alpha_0 + \alpha_1 + \alpha_2$.
The Weyl group $W$ of $a_2^{(1)}$ is generated by the
reflections $w_\alpha, \alpha\in \Delta^{re}_+$,
with its two actions

\[
  w_\alpha(\lambda)
= \lambda - \alpha ( \lambda, \alpha^\vee )
\;,\;\;\;\;
  w_\alpha\cdot\lambda
= w_\alpha( \lambda +\rho) - \rho
\;,
\]
 where $\rho$ is given below.
The coroots of $a_2^{(1)}$ are defined by 
$ \alpha^\vee = 2 \alpha /(\alpha,\alpha) $.  The basis of  fundamental
weights $\{\omega_i\} $ is dual to that of  the  simple coroots:
$ (\omega_i , \alpha_j^\vee) = \delta_{ij} $.
Any weight $\lambda$ of $a_2^{(1)}$ can be decomposed as
\[
  \lambda 
= \sum_{i=0}^2 \lambda_i \omega_i
\equiv [\lambda_0,\lambda_1,\lambda_2]
\;,
\]
the numbers $\lambda_i$ being called the Dynkin labels. The level $k$
of the weight is the sum of the Dynkin labels: 
$k= \sum_{i=0}^2\lambda_i$ (each fundamental weight has level 1).  The
special weight with all three Dynkin labels equal to 1 (which thus has
level 3) will be denoted by $\rho=\rho^\vee=[1,1,1]$. Any weight
$\lambda$ of $a_2^{(1)}$ can be projected onto a weight $\bar\lambda$
of $a_2$ by setting the coefficient of $\omega_0$ to zero. 

With our normalisation, the parametrisation of $c,h$ and $w$ in
\cite{vddv} is 
\be
\begin{array}{rcl}
 c &=& 50 - 24 t - 24 /t\;,  \\
 h &=& {\displaystyle 
       \frac{1}{2t}| \blp |^2 + \frac{c-2}{24}\;, }\\
 w &=& {\displaystyle \frac{1}{27 t^{3/2}} }
       ( \blp,\omega_1)( \blp,\omega_2)( \blp,\omega_1-\omega_2)
\;,
\end{array}
\label{eq:param}
\ee
where, as explained, $\lambda+\rho$ is a weight of $a_2^{(1)}$ of
level $k+3=t>0$, and $\blp$ is its projection onto
the weight space of $a_2$.  

It is explained in \cite{vddv}  how the structure of a $\WA$
highest-weight representation is governed by three particular
subgroups of the Weyl group of $\slt$ determined by $\lambda$:
$W_\lambda, W^0_\lambda$ and $W^r_\lambda$.
The group $W_\lambda$ is the subgroup of $W$ generated by 
$w_\alpha$ for all $\alpha \in \Delta^{re}_{\lambda,+}$, where
\[
  \Delta^{re}_{\lambda,+} 
= \{
  \alpha \in \Delta^{re}_+ | (\lambda,\alpha^\vee)\in Z
  \}
\;.
\]
We denote the simple roots of $\Delta^{re}_\lambda$ by $\hat\alpha_i$,
and the corresponding  Weyl groups elements by $\hs i$. 
Now consider the orbit $W_\lambda\cdot\lambda$, which will contain a
unique dominant weight $\lambda_{dom}$, characterised by
$(\lambda_{dom}+\rho,\alpha_i^\vee)\geq 0$ for all $i$.
$W^r_\lambda$ is generated by the $\hs i$ which
preserve $h$ and $w$ for each weight in this orbit
and $W^0_\lambda$ is generated by the $\hs i$ which preserve the
dominant weight itself.

The structure of the $\WA$ highest-weight representations
corresponding to any weight in the $W_\lambda$ orbit of $\lambda$ is
entirely specified by the double coset 
\be
  W^r_\lambda \backslash W_\lambda / W^0_\lambda
\label{eq:dcst}
\ee
and the associated Kazhdan-Lusztig polynomials, in the manner
explained in \cite{vddv} and reviewed below.  In particular, 
Conjecture~1 of Ref.~\cite{vddv}  enables us to find the composition
series for any $\WA$ highest-weight Verma module representation,
and consequently the character of any irreducible highest-weight
representation.

The composition series of a highest-weight representation $M_0$ is a 
series of irreducible highest-weight modules $L_i$ such that 
$L_i = M_i / M_{i+1}$. The set $\{L_i\}$ obtained in this way is
independent of the actual series of modules 
$M_0 \supset M_1 \supset M_2 \ldots$.
We can then write, as vector spaces, 
\be
  M_0 = \,L_0 +a_1\, L_1 +a_2\, L_2 + \ldots
\;.
\label{eq:cs}
\ee
where $a_i$ is the multiplicity of $L_i$ in $M_0$. 
According to de Vos and van Driel, if $M_0$ is associated to a weight
$w \cdot \lambda$ where $\lambda$ is dominant, then the only
irreducible representations which occur in  (\ref{eq:cs}) are those
associated to weights $w' \cdot \lambda$ where  
$w',w\in W^r_{\lambda}\backslash W_\lambda/ W^0_{\lambda}$ and
$w'\geq w$ (in terms of the Bruhat ordering, meaning that $w$ can be
obtained from the decomposition of $w'$ in terms of simple Weyl
reflections by dropping one or more factors).
The multiplicity of $L_{w'\cdot\lambda}$ in $M_{w\cdot\lambda}$ is
given by the Kazhdan-Lusztig polynomial $P_{\bar w,\bar w'}(1)$, where
$\bar w$ stands for the maximal representative of $w$ in
$W_{\lambda}$. As a result, the character of $M_0$ is given as
\be
  \ch{M_{w\cdot\lambda}}
= {\rm Tr}_{M_{w\cdot\lambda}}( q^{L_0} )
= \sum_{{ w'\geq w } \atop 
        { w',w \in   W^r_{\lambda}\backslash W_\lambda/W^0_{\lambda}} }
  \!\! P_{\bar w,\bar w'}(1)
  \;\; \ch{L_{w' \cdot \lambda}}
\;.
\label{eq:char1}
\ee
(where $L_0$ appearing in the power of $q$ is the Virasoro zero mode,
and we have left out the usual factor $q^{-c/24}$.) However, up to a
prefactor, the character of a Verma module is independent of the
weights $h,w$,  
\[
  \ch{M_{h,w}}
= q^{h} \prod_{i>0} (1 - q^i)^{-2}
\;,
\]
and it is now simple to calculate the character of the irreducible
representations $L_{w \cdot \lambda}$ by inverting (\ref{eq:char1}) as
\be
  \ch{L_{w\cdot\lambda}}
= \sum_{{ w'\geq w } \atop 
        { w',w \in   W^r_{\lambda}\backslash W_\lambda/W^0_{\lambda}} }
  \sum_{ x \in [w] }
  \!    Q_{\bar w,x}(1) (-1)^{l(\bar w) + l(x)}
  \;\;  \ch{M_{w'\cdot\lambda}}
\;.
\label{eq:char2}
\ee
it where $Q$ stands for the inverse  Kazhdan-Lusztig polynomial (see
below). The polynomials $P_{w,w'}(q)$ can be calculated recursively
{}from  
\be
  P_{x,sy}
= q^{1-b}P_{xs,y}+q^bP_{x,y}-q\sum_{x\leq z<y\atop z s<z}
  P_{x,z}\;{\tilde P}_{z,y}
\;,
\label{eq:recur}
\ee
using $P_{x,e}=1$.  Here $s$ is a simple Weyl reflection, $x,y$ are
elements of $W_{\Lambda}$.  The number $b$ is 1  if $xs<x$ and 0
otherwise. Finally, ${\tilde P}_{x,z}$ is the term of degree
$(1/2)( l(x) - l(z) - 1)$ in $P_{x,z}$ and $l(x)$ is the smallest
number of simple Weyl reflections with which one may produce the
element $x$. 
The polynomials $Q_{x,y}(q)$ then satisfy
\[
  \sum_{x \leq z \leq y}
  P_{x,z}(q) Q_{z,y}(q) (-1)^{ l(z) + l(y) }
= \delta_{x,y}
\;.
\]

When $t= p/p'$ is rational and $W_\lambda\equiv W$, de~Vos and
van~Driel give the general form of a dominant weight $\lambda$ and the
subgroups $W^0_\lambda$ and $W^r_\lambda$. In this 
case, 
\be
  \lambda + \rho = w( \Lambda_+ - t \Lambda_- )
\;,\;\;
  \Lambda^+ = \sum_{i=0}^2 \Lambda^+_i \omega_i
\;,\;\;
  \Lambda^- = \sum_{i=0}^2 \Lambda^-_i \omega_i
\;,
\label{eq:lplm}
\ee
where $w$ is an element of the finite Weyl group generated by
$w_{\alpha_1}, w_{\alpha_2}$, $\Lambda^+$ is a weight of level $p$ and
$\Lambda^-$ is a weight of level $p'-1$, and where $\Lambda^+_i,
\Lambda^-_i$ are non-negative integers. Then $W^0_\lambda$ is
generated by the $\hs i$ for which $\Lambda^+_i$ are zero, and
$W^r_\lambda$ is generated by the $\hs i$ for which $\Lambda^-_i,
i=1,2$ are zero. With this parametrisation, the simple roots of
$\Delta^{re}_{\Lambda,+}$ are 
\be
\hat\alpha_i = w(\alpha_i) + \Lambda^-_i\delta.
\label{eq:alhat}
\ee
This being the situation of interest to us, from now on we shall
denote the groups $W^0_\lambda$ and $W^r_\lambda$ by  $W_{\Lambda^+}$
and  $W_{\bar\Lambda^-}$ respectively. 

The minimal models of $\WA$  correspond to admissible weights of
$a_2^{(1)}$ and  have $t=p/p'$ with $p,p'$ coprime, $p,p' \geq 3$.
Their representations are of the form (\ref{eq:lplm}) and they are
denoted by $[\Lambda^+_1 \Lambda^+_2; \Lambda^-_1 \Lambda^-_2]$ with
the labels satisfying $\Lambda^\pm_i \geq 1$, 
$\Lambda^+_1+\Lambda^+_2< p$, $\Lambda^-_1+\Lambda^-_2 < p'$.  
The vacuum representation is type $[11;11]$. The field
corresponding to the weight $[ab;cd]$ will be written $\PHI abcd$.

\subsection{
Structure of $\WA$ vacuum representations and the quotient $\cF $}
\label{sec:kl}

We are primarily interested in the space of fields spanned by normal
ordered combinations of $L$, $W$ and their derivatives.
These fields are in one-to-one correspondence with the states in the
quotient of the $h=w=0$ Verma module by the space generated by
$L_{-1}\vac$ and $W_{-1}\vac$.\footnote{
Notice that $L_{-1}\vac=W_{-1}\vac=0$ forces $W_{-2}\vac=0$ so that
the nonvanishing actions of the algebra generators on the vacuum are
$L_{-2-n}\vac\sim \partial^{n}L$ and  $W_{-3-n}\vac\sim \partial^{n}W$
for $n=0,1,\ldots$}
We shall call this quotient space $\cF$. Using the results of
\cite{vddv}, we can systematically search for all models for which
there are singular vectors in $\cF$.  This is the subject of the
present section.  In a second step, we will examine whether the
singular vectors in thoses models do in fact give conserved densities
for any integrable perturbations. This will be done in the following
section. 

For a generic value of $c$, that is, for $t$ irrational, 
$\cF$ corresponds to the irreducible highest-weight representation
with $h=w=0$ and
contains no singular vectors. If $t$ is rational, however, then
$\cF$ may contain some singular vectors, and it is only these
singular vectors which can serve as our singular conserved densities.
The simplest way to determine if there are singular vectors in 
$\cF$ by computing characters. The character of $\cF$ is
\be
  \ch\cF 
= {\rm Tr}_\cF ( q^{L_0} )
= (1-q^2)^{-1} \prod_{i>2} (1 - q^i)^{-2}
\;.
\label{eq:fchar}
\ee
The irreducible vacuum module $L_{vac}$ is the quotient of $\cF$ by
its maximal submodule, which is a highest-weight submodule of $\cF$
of which the highest-weight states are singular vectors in $\cF$. 
Consequently, if the character  $\ch L_{vac}$ is found to be identical
to $\ch\cF $, it means that there are no singular vectors in $\cF$. 
 Consequently, any difference between the two characters signals
the presence of singular vectors in $\cF$, which may then be readily
identified. The formalism in \cite{vddv} allows one to calculate the
character of any highest-weight representation, and in particular the
vacuum representation, from the Kazhdan--Lusztig polynomials
associated to the coset (\ref{eq:dcst}). 

We recall that the vacuum representation is type $[11;11]$,
\be
  \lambda_{vac} + \rho
= [p-2,1,1]-\frac{p}{p'} [p'-3,1,1]
\;.
\label{eq:can}
\ee
Let us restrict to rational $t>0$; then without loss of generality, we
take $p \geq p'$, $p,p'$ coprime. We then consider the three cases,
$p'=1$, $p'=2$ and $p'\geq 3$ in turn. 

\subsubsection{$\cF$ for $p'=1$}

When $p'=1$, the Weyl group element $w$ in (\ref{eq:lplm}) plays no
role since the action of a finite Weyl reflection on the finite simple
roots give an equivalent basis of simple roots, and we may take
$\Lambda^- = 0$ and  $\hat\alpha_i = \alpha_i $. The analysis splits
further into the three subcases: 
$t=p=1$,
$t=p=2$ and
$t=p\geq 3$.

\begin{itemize}

\item{$p=1$}

For $p=1$, $\lambda_{vac}$ is dominant, so that in terms
of the parametrisation  (\ref{eq:lplm}) this corresponds to 
\[
  \Lambda^+ = [1,0,0]
\;,\;\;
  \Lambda^- = [0,0,0]
\;,
\]
and $W_{\Lambda^+} = W_{{\bar \Lambda}^-} = a_2$. This is the first
case in table~2 of \cite{vddv}; the corresponding embedding diagram
is given in their table~5. This model has central charge 2; since it
has an explicitly unitary representation in terms of two free bosons,
it is clear that there are no singular vectors in $\cF$.  

\item{$p=2$}

For $p=2$, $\lambda_{vac} $ is not dominant; the dominant weight
$\lambda$ in the orbit $W_{\lambda_{vac}} \cdot \lambda_{vac} $ is 
\[
  \lambda
= \hs 1 \hs 2 \hs 1 \cdot  \lambda_{vac} 
\;,\;\;
  \lpp = [0,1,1]
\;.
\]
Here $W_{\Lambda^+}  = a_1$, $W_{{\bar \Lambda}^-} = a_2$ and
$\lambda$ also corresponds to the $\WA$ vacuum representation of this
model. The embedding structure has been worked out explicitly in
\cite{vddv}, corresponding to the fourth case in their table 2, with
the embedding diagram given in their table 8. 

To see whether there are singular vectors in $\cal F$, we need to
calculate $\ch L_{vac}$ using the KL polynomials given in \cite{vddv}
and compare it with  $\ch {\cal F}$. However, since de Vos and van
Driel do not give a general formula for the KL polynomials in  this
case, we cannot rule out singular vectors in $\cF$  at levels below
those given in  \cite{vddv}, table 8, although we believe that $\cF$
has no singular vectors at all.

\item{$p\geq 3$}

For $p\geq2$ $\lambda_{vac} $ is not dominant;  the
dominant weight in its  Weyl orbit is 
\[
 \lambda 
= \hs 0\hs 1 \hs 2 \hs 1 \cdot  \lambda_{vac}
\;,\;\;
  \lpp = [p-2,1,1]
\;,
\]
which corresponds to a $\WA$ representation with $h=2-p, w=0$, with 
the $\WA$ vacuum representation the first singular descendant of 
this highest weight. Thus we get \hbox{$W_{\Lambda^+}=1$},
\hbox{$W_{{\bar\Lambda}^-}=a_2$} and the embedding structure is also
treated  explicitly in \cite{vddv}, corresponding to the fifth case in
their table 2,  with the embedding diagram given in table 9. 

There are no singular vectors in $\cF$ here since the
character of the irreducible vacuum representations $L_{vac}$ is
identical to that of $\cF$.

\end{itemize}

\subsubsection{$\cF$ for $p'=2$}
\label{ssec:lm}

For $p'=2$, $\lambda_{vac}+\rho = [3p/2-2,1-p/2,1-p/2]$ and 
we can choose the simple roots $\hat\alpha_i$ as
\[
  \hat\alpha_0 = \alpha_1 + \alpha_2
\;,\;\;
  \hat\alpha_1 = -\alpha_1 + \delta
\;,\;
  \hat\alpha_2 = -\alpha_2 + \delta
\;,\]
$\lambda_{vac}$ is not dominant since 
$  (\lambda_{vac}+\rho,\hat\alpha_1) 
 = (\lambda_{vac}+\rho,\hat\alpha_2) =-1$ and the dominant weight in  
$W_{\lambda_{vac}}\cdot \lambda_{vac}$ is
\[
 \lambda
= \hs 0 \cdot \lambda_{vac}
\;,\;\;\;\;
 \lpp= [2-\frac p2,\frac p2-1,\frac p2-1]
\;.
\]
Since $(\lpp,\hat\alpha_1) = (\lpp,\hat\alpha_2) =1$ and
$(\lpp,\hat\alpha_0) = p-2$ are all positive, $\lambda$ is
dominant. In this case $\lambda$  also corresponds to the vacuum
representation. $\lpp$ can also be written in the form (\ref{eq:lplm})
\[
 \lpp
= w_{\alpha_1} \left(
    [1,1,p-2]
    \;-\;
    \frac p2 [0,1,0]
              \right)
\;,\;\;
\]
We find $W_{\Lambda^+} = 1$, $W_{{\bar\Lambda}^-} = a_1$. The embedding
structure is described in  table 12 of \cite{vddv}.

If we compute the character $\ch L_{vac}$ in this case, we find that
it is not equal to $\ch \cF $,
\be
  \ch L_{vac}
= \ch \cF  \,
  \left( 1 - 2 q^{3p-3} + \ldots \right)
\;,
\label{eq;char3}
\ee
and consequently there is at least one singular vector in $\cF$ at
level $3p-3$. To be more  explicit, if we denote by $M(i)$ and $L(i)$
respectively the highest-weight Verma module and irreducible
highest-weight module corresponding to the node $(i)$ of table 12 in
\cite{vddv}, at the level of vector spaces,  we find that
\begin{eqnarray*}
  M(1)
&=&
  L(1) + L(2) + L(3) + L(4) + L(5) + L(6) + L(7) + L(8)
\\&&+ 2 L(9) + L(10) + 2 L(11) + L(12) + \ldots
\\
  M(2)
&=&
  L(2) + L(4) + L(5) + L(6) + L(8) +
  L(9) + L(10) + L(11) + L(12) + \ldots
\\
  M(3)
&=&
  L(3) + L(4) + L(6) + L(7) + L(8) +
  L(9) + L(10) + L(11) + L(12) + \ldots
\\
\end{eqnarray*}
However, we know that for each distinct $L(i)$ appearing in the sum of
$M(1)$ there is a unique highest-weight vector in $M(1)$
\cite{bajnok}, and consequently the highest weight of type $L(9)$
appearing in $M(2)$ must be identical to that in $M(3)$, and we find 
\begin{eqnarray*}
  M(2) \cap M(3)
&\geq&
  L(4) + L(6) + L(8) + L(9) + L(10) + L(11) + L(12) 
\\
  M(2) \cup M(3)
&\leq&
  L(2) + L(3) + L(4) + L(5) + L(6) + L(7) +L(8)
\\&&+ L(9) + L(10) + L(11) + L(12)
\end{eqnarray*}
The two singular vectors at level one in this vacuum module
corresponds to the nodes 2 and 3. $\cF $ is then given by $ M(1)/(
M(2)\cup M(3))$ and we find that 
\[
  \cF  
= M(1)  \Big/ ( M(2) \cup M(3) ) \leq  L(1) + L(9) +  L(11) 
\]
We see that there are indeed singular vectors of type $(9)$ and $(11)$
in $\cF$, that is singular vectors of types $[11;14]$ and $[11;41]$
of weight $3p-3$. These are candidates for singular conserved
currents. 

\subsubsection{$\cF$ for $p' \geq 3$}
\label{ssec:min}

For $p' \geq 3$, $\lambda_{vac}$ is dominant, $W_{\Lambda^+} = 1$,
$W_{{\bar\Lambda}^-} = 1$ and
\[
  \hat\alpha_0 = \alpha_0 + (p'-3) \delta
\;,\;\;
  \hat\alpha_1 = \alpha_1 + \delta
\;,\;
  \hat\alpha_2 = \alpha_2 + \delta
\;.
\]
The embedding structure is given in table 13 of \cite{vddv}. The
vacuum representation has highest-weight vectors of types (1), (3) and
(5) (using the notation in table 13 of \cite{vddv}), at levels 1,1 and
$(p-2)(p'-2)$ respectively. Any other highest-weight vector in the
vacuum Verma module is conjectured to be a descendant of these three.
Repeating the above  calculation for $\cF$, we find a singular
vector at level $(p-2)(p'-2)$ as our candidate conserved current, and
it transforms in the $[1,1;p'-1,p'-1]$ representation. 

\subsection{Searching for integrable perturbations}
\label{sec:search}

We have only found two series of models for which there are singular
vectors in $\cF$, namely the $p'=2$ models and the minimal models. We
now systematically search for perturbing fields for which the
conserved densities could be singular vectors in $\cF$. An important
difference between the Virasoro and a general $W$ case must be
stressed from the start.  In the Virasoro case, we can solve the
descent equations and find the o.p.e. for any three fields with
arbitrary weights $h_i$ (\ref{eq:ope2}), and taking $\Delta h =2$ is
enough to conclude that the single pole is a total derivative
coefficient. This is no longer true for the $W_3$ algebra. The general
o.p.e. of two $\WA$--primary fields is 
\be
  \phi_{h_1,w_1}(z)   \vec{h_2,w_2}  \;
\propto
  z^{h_3 - h_1 - h_2} \Big[   
  1 
+ z \big(
  \frac{ h_1 + h_3 - h_2 }{ 2 h_3 } L_{-1}
  + 
  \mu \left[ W_{-1} - \frac{3 w_3}{2 h_3} L_{-1} \right] 
  \big)
+ \ldots  \Big]\vec{h_3,w_3}
\;, 
\label{eq:wope4} 
\ee
where $\mu$ is an {\it a priori} undetermined constant. It is only by
enforcing further constraints, such as the decoupling of null vectors,
that $\mu$ can be determined. Furthermore, we see that taking $\Delta
h=2$ is not sufficient to make the residue of the pole a total
derivative, and that we require in addition that either $\mu = 0$ or 
$\left[ W_{-1} - \frac{3 w_3}{2 h_3} L_{-1} \right] \vec{h_3,w_3} $ 
is a null state.

As in the Virasoro case, we first consider which fields may occur in
the o.p.e. 
\be
 \PHI abcd \;\;\PHI rstu \sim \sum \PHI{r'}{s'}{t'}{u'}
\;.
\label{eq:wope2}
\ee
where $\PHI abcd $ plays the role of the (singular) conserved density
and $\PHI rstu$ that of the perturbation.

We first ensure that the coefficient of $\PHI rstu$ on the r.h.s. of
(\ref{eq:wope2}) is zero (i.e., to ensure the decoupling of the
singular vector $\PHI abcd$ from the physical states); this will
restrict our choices of $[rs;tu]$. Secondly, since $\PHI abcd$ is
required to be singular, hence a polynomial in $L$ and $W$, we may
restrict the sum in the r.h.s. of (\ref{eq:wope2}) to descendants of
$\PHI rstu$.  

We consider in turn the minimal models and then the $p'=2$ case.

\subsubsection{Integrable perturbations for $p'\geq 3$}

For a $(p,p')$ $\WA$ minimal model, the candidate conserved current
was shown in sect. \ref{ssec:min}  to be necessarily of the form 
$\PHI 11{p'-1}{p'-1}$.  We now need to determine those  values of $p'$
for which this is indeed conserved under some perturbations. 
To ensure that $\PHI rstu$ does not appear in the r.h.s.\ of
(\ref{eq:wope2}), $\PHI rstu$ must be taken in the set of minimal
representations, and we choose to set the  null vectors at levels $rt$
and $su$ to zero. This leaves a single $\WA$ primary descendant at
level $(p-r-s)(p'-t-u)$, of type $[rs;p'-u,p'-t]$, in the r.h.s of
(\ref{eq:wope2}). To check the conservation of 
$\int\PHI 11{p'-1}{p'-1}$ under some perturbation $\int\PHI rstu$, we
consider the o.p.e. (\ref{eq:wope4}) and find 
\be
  \Delta h
= (p'-t-u)(r+s-2) + (p-r-s)(t+u-2) + (r+s-2)(t+u-2)
\geq 0
\;.
\label{eq:wope}
\ee
If $\Delta h = 2$ and $\mu = 0$, the conservation of 
$\int\PHI 11{p'-1}{p'-1}$ is guaranteed. Setting $\Delta h = 2$, and 
choosing $p \geq {p'}$, $s \geq r$, we get  three possible solutions: 
\be
\begin{array}{r|lllll}
    & {p'} & r & s & t & u \\ \hline
(a) &  4   & 1 & 2 & 1 & 1 \\
(b) &  3   & 1 & 3 & 1 & 1 \\
(c) &  3   & 2 & 2 & 1 & 1
\end{array}
\ee
The coefficient $\mu$ in (\ref{eq:wope}), appropriate to each case, is
calculated in appendix \ref{app:wope}:  $\mu$ does not vanish in the
cases  (a) and (b), but it does in case (c).

So, we have found an integrable perturbation which lies in the Kac
table of a generic $(p,p')$ minimal model and this is  the $[22;11]$
field. In this case,  there is a conserved current transforming in the
$[11;22]$ representation. This representation is in the vacuum module
whenever $p'=3$. For this value of $p'$, the current has weight $p-2$,
and so the corresponding conserved charge has weight $p-3$; since $p$
must be coprime to 3, the weights of the charges in the different
$(p,3)$ models are all coprime to 3. Thus each exponent%
\footnote{The exponents of $a_2^{(1)}$ are the positive integers
coprime to 3}
of $a_2^{(1)}$arises as the weight of a singular vector conserved
charge for some value of $c$. By interchanging the role of $p$ and
$p'$, we also conclude that $[11;22]$ is an integrable perturbation.
The duality argument does not provide any further solutions since
$[22;11]$ and $[11;22]$ are dual to each other (cf. section
\ref{ssec:a2}).   

\subsubsection{Integrable perturbations for $p'=2$}

The analysis of $\cF$ for $p'=2$ shows that singular vectors can
only be of type $[11;14]$ or $[11;41]$.  We now have to determine
under which perturbations these can be conserved densities. Attempting
to repeat the previous argument, we face an immediate problem: since 
the $(p,2)$ models are not minimal, there are no obvious
candidates for the representations $[rs;tu]$ (corresponding to the
perturbing field) which are compatible with the vanishing of all the
null vectors in the vacuum module. To determine this would be a
complete calculation of the so-called Zhu's algebra \cite{zhu,mecmp}
for the irreducible vacuum module, which we do not wish to address
here. However for small values of $p$, we have examined the
restrictions obtained by requiring the null vectors of type $[11;14]$
and $[11;41]$ of weight $3p-3$ to decouple from the physical
correlation functions. These calculations are detailed in appendix
\ref{app:zhu}, and indicate that the representations of interest are
of type $[rs;11]$ where one of the following holds: 
$ r = 1,2 \ldots p-1$,
$ s = 1,2 \ldots p-1$,
$ 2(r+s) = p+2,p+4 \ldots 3p-2$.

We shall restrict our attention to the Verma modules with both $r$ and
$s$ integer, which are of the same type as the vacuum, (cf.
\cite{vddv}, table 12). Correspondingly, if we set to zero the
singular vectors at levels $r$ and $s$, we find two remaining singular
vectors at levels $3p-2r-s$ and $3p-r-2s$, which give for $\Delta h$
in (\ref{eq:wope}), 
\[
  \Delta h = \cases{ 2s + r - 3 & \cr s + 2r - 3 & \cr }
\]
Putting $\Delta h = 2$, we are lead to the following four possible
integrable perturbations:
\be
\{\;
  [12;11] \;,\;\;
  [21;11] \;,\;\;
  [13;11] \;,\;\;
  [31;11]
\;\}\;.
\label{eq:plist}
\ee

Let us restrict attention to the singular vector in $\cF$ of type
$[11;14]$. We now examine the o.p.e.\ of this field with each field in
(\ref{eq:plist}). We stress that the o.p.e. of $[11;14]$ with any of
the four fields in the above list can in principle produce a field
with $ \Delta h=2$. However, this triple coupling is not necessarily
allowed: we need to  show  that the third field is a genuine singular
descendant of the field whose o.p.e. with  $[11;14]$ is considered.

\subsubsection{$[11;14] \times [13;11]$ }

In this case there are two singular-vector descendants of $[13;11]$ to
which we may couple, viz.\ $[13;14]$ and $[31;14]$. Of these the first
is at level $3p-7$ which gives $\Delta h = 4$; this is thus ruled out.
In the second case, there is no null vector at level 1, and repeating
the arguments of appendix \ref{sec:wacoeffs}, we see that $\mu$ does
not vanish for this perturbation. Hence the $[11;14]$ field is not a
conserved quantity for the $[13;11]$ perturbation. 

\subsubsection{$[11;14]\times [31;11]$}

As in the previous subsection, there are two singular-vector
descendants of $[31;11]$ to which we may couple, and we find that
neither of these is satisfactory, and hence the $[11;14]$ field is not
a conserved quantity for the $[31;11]$ perturbation.

\subsubsection{$[11;14] \times [21;11]$}

In this case, we know from the analysis of \cite{BPT} that the
presence of null vectors at levels $1$ and $2$ in the $[12;11]$ Verma
module imply that the only possible $\WA$ representations which may
occur in the o.p.e. $[11;14] \times [21;11]$ are $[21;14]$, $[10;14]$
and $[02;14]$. Of these, only the first is a descendant of the
$[12;11]$ highest weight, and this is exactly the singular vector at
level $3p-4$. However, this gives $\Delta h$ = 1 and we do not find a
conserved current.  

\subsubsection{$[11;14] \times [12;11]$  }

As before, the null vectors at levels $1$ and $2$ in the $[12;11]$
Verma module imply that the only possible $\WA$ representations which
may occur in the o.p.e.\  $[11;14] \times [12;11]$ are $[12;14]$,
$[01;14]$ and $[20;14]$. Of these, only the first is a descendant of
the $[12;11]$ highest weight, and this is exactly the singular vector
at level $3p-5$. Since this is of type $[12;14]$, there is a singular
vector at level 1, and so the coefficient $\mu$ in (\ref{eq:wope}) can
be set to zero. Thus we find that the singular vector in $\cF$ of type
$[11;14]$ is a conserved density for a perturbation of type $[12;11]$. 

The general analysis of Feigin and Frenkel \cite{FF} also shows that
the $[11;12]$ and $[11;14]$ perturbations should also be integrable,
corresponding to the dual pair of $g_2^{(1)}$ and $d_4^{(3)}$ affine
Toda theories%
\footnote{This can also be seen from a simpler analysis, based on the
Zamolodchikov counting argument, presented in \cite{Vays}.
}.
We find that in the $(p,2)$ models, the singular vector at level
$3p-3$ in $\cF$ is a  conserved quantity for this perturbation,
leading to conserved charges of weight $\Delta = 3p - 4$, $p$ odd,
which gives the series 
\[
\begin{array}{c|ccccccc}
p      & 3 & 5 & 7 & 9 & 11 & 13 & \cdots\\ \hline
\Delta & 5 & 11&17 & 21& 29 & 35 & \cdots
\end{array}
\]
These numbers form a subset of the exponents of $d_4^{(3)}$ or
$g_2^{(1)}$ , which are the positive integers coprime to 6.

\subsection{Discussion}

We have examined the quotient space  $\cF$  and have identified the
cases such a field may be a null field. These fall into two classes,
namely the $(p,2)$ models and the $(p,p')$ models for $p' \geq 3$. In
the first case, we have found that the $[12;11]$ and $[21;11]$
representations are integrable with conserved quantities of weight
$3p-4$ for $p$ coprime to 3, transforming in the $[11;14]$ and
$[11;41]$ representations respectively.  This leads to  a set of
conserved charges with weights that form subset of the exponents of
$g_2^{(1)}$. By duality (cf. appendix A), $[11;14]$ and $[11;41]$ are
also integrable perturbations. For the minimal models, we have  found
that the $[22;11]$ perturbation is integrable for $p'=3$ with
conserved densities of weight $p-2$ transforming in the $[11;22]$
representation. The conserved quantities have weights that take all
the values of the exponents of $a_2^{(1)}$. Thus our analysis exhausts
the perturbations which are known to be integrable from the results of
\cite{FF}.  

As for the Virasoro algebra, we could also consider conserved
densities with $\Delta h \geq 3$, but we have not investigated these.  
But again, only sporadic solutions, if any, are to be expected in
those cases: the conditions required to have a simple pole with total
derivative residue are much more stringent here than in the Virasoro
case. Two more situations, which did not occur for the Virasoro
algebra, have been unexplored. Firstly, we have only considered triple
couplings of primary fields. However, the leading non-zero term in an
o.p.e.\ of $\WA$ primary fields (\ref{eq:wope4}) need not be a
$\WA$--primary field. A simple example is provided by the three-state
Potts model which has  $c=4/5$ and a highest-weight representation
with $\{h=2/5,w=0\}$, whose corresponding field is written
$\phi_{2/5}$. Then the leading term in the o.p.e.\ of two $\phi_{2/5}$
fields is  
\[
  \phi_{2/5}(z) \vec{2/5}
\propto 
  z^{3/5} W_{-1} \vec{2/5} + \ldots
\;,
\]
which is a descendant field. Again, we have not investigated this
case. The second possibility which has been ignored, related to the
first, is that descendant fields could  provide integrable
perturbations. Again the three-state Potts model provides an
illustration of this situation: the chiral three-state Potts model is 
an integrable perturbation of the critical three-state Potts model by
the non-$\WA$ primary field $W_{-1}\phi_{2/5}$. For this particular
case, we have investigated whether this field, which corresponds to
the Virasoro--highest-weight state $W_{-1} \vec{11;22}$, is an
integrable perturbation for other values of $c$ than $c=4/5$, and we
find that it is not. Consequently we do not expect to find any new
series of integrable perturbations by descendant fields.  

\section{The $\WB$ algebra and its representations}
\label{sec:wb}

\subsection{Introduction}
\def\ds{\displaystyle}
The $\WB$ algebra was first described in \cite{wc2}. Its generators
are $L_m$ and $W_m$, which satisfy the usual Virasoro relations and
\[
\begin{array}{rl}
 ~[L_m,W_n]
&=\ds
 (3m-n)W_{m+n} \\
 ~[W_m,W_n]
&=\ds
 \gamma^2
 \,\frac{ c }{ 4}
 \,\frac{ m(m^2-1)(m^2-4)(m^2-9) }{ 7! }
 \,\delta_{m+n}
\\[3mm]
&\!\!\!\!\!\!\!\!\!\!\!\!\!\!\!\!\!\!\!\!\!\!\!\!\!\!\!\!\!\!
+\,\ds
  \gamma^2
  \,\frac{(m-n)}{1680}
  (3n^4-2mn^3+4m^2n^2-39n^2-2m^3n+20nm-39m^2+108+3m^4)
  \,L_{m+n}
\\[3mm]
&+\,\ds
  \frac 1{36} (m-n) (m^2 - mn + n^2 - 7 )
  \,W_{m+n}
\\[3mm]
&+\,\ds
  \frac{7}{324}
  \frac { (m - n )( m^2 - mn + n^2 - 7 ) (7c + 68 ) (2c - 1 )}
        { ( c + 24 )(c^2 - 172c + 196 )}
  \,\Lambda_{m+n}
\\[3mm]
&+\,\ds
  \frac2{9}
  \frac { (m - n )(72 c + 13 )}
        { (c + 24 )(c^2 - 172c + 196 )}
  \,\Delta_{m+n}
\\[3mm]
&+\,\ds
  \frac{1}{180}
  \frac { (m - n )(19c - 524 )(5c + 22 )}
        { (c + 24 )(c^2 - 172c + 196 )}
  \,\Gamma_{m+n}
\\[3mm]
&+\,\ds
  \frac{14}3
  \frac { (m - n )}{ c+24}
  \,\Xi_{m+n}
\end{array}
\]
where
\[
\begin{array}{rcl}
 \gamma^2 &=&
 \frac{\ds (5c+22)(7c+68)(2c-1)}{\ds 54(c+24)(c^2 - 172c + 196 )}
\\
 \vec\Lambda &=& \left(
  L_{-2}L_{-2} - \frac35L_{-4}
                  \right)\vac
\\
  \vec\Delta  &=& \left(
  {L_{-2} L_{-2} L_{-2}}
- \frac 23   {L_{-6}}
- \frac {19}{15} {L_{-4} L_{-2}}
- \frac {1}3   {L_{-3} L_{-3}}
                \right) \vac
\\
  \vec\Gamma  &=& \left(
  \frac {10}{63} {L_{-6}}
+ \frac 4{9}   {L_{-4} L_{-2}}
- \frac 5{18}  {L_{-3} L_{-3}}
                \right) \vac
\\
  \vec\Xi     &=& \left(
  {L_{-2} W_{-4}}
- \frac {1}3  {W_{-6}}
                \right)\vac
\end{array}
\]
This is related to the usual normalisation by
\[
W = \gamma \; W_{standard}
\;.
\]

\subsection{Quantum Hamiltonian reduction and highest-weight representations}

The $\WB$ algebra arise naturally from the quantum Hamiltonian
reduction of $c_2^{(1)}$, as has been performed in \cite{IT}.
Let us recall some properties of the  $c_2^{(1)}$ algebra. The
fundamental weights of $c_2^{(1)}$ are $\omega_0$, $\omega_1$ and
$\omega_2$ of level 1, $\rho = [1,1,1]$, and we now define
$\rho^\vee = [1,2,1]$. The simple roots are $\alpha_0$, $\alpha_1$ and
$\alpha_2$ with 
\[
  |\alpha_0|^2 = |\alpha_2|^2 = 2
\;,\;\;\;
  |\alpha_1|^2 = 1
\;,\;\;\;
  (\alpha_i,\alpha_j) = -1,\;\; i \neq j
\;.
\]
The coroots are $\alpha^\vee = 2 \alpha / |\alpha|^2$ and the
fundamental coweights are $\omega^\vee_i = 2\omega_i/|\alpha_i|^2$,
which satisfy $(\alpha_i,\omega_j^\vee) = \delta_{ij}$ and are the
fundamental weights of $d_3^{(2)}$.

The highest-weight representations of $\WB$ are labelled by the
central charge $c$, and by $\{h,w\}$, the eigenvalues of $L_0$ and
$W_0$. As before, each highest--weight representation arises from the
quantum Hamiltonian reduction of some $c_2^{(1)}$ representation, and
we may parametrise $c,h$ and $w$ in terms of the corresponding
$c_2^{(1)}$ weight $\lambda$ of level $ t-3$ by 
\[
\begin{array}{rcl}
    \lpp
&=& (t-x-y)\omega_0 + x \omega_1 + y \omega_2
\;,\\

 c &=& 86 - 30 t^{-1}  - 60 t
\;, \\

 h &=& {\displaystyle \frac{1}{4t} \left(  x^2 + 2 x y + 2 y^2 \right)
       + \frac{c-2}{24} }
\;, \\[4mm]

 w &=& {{\left( \begin{array}{l}
         (2 t^2 - 1)( 14 t (x^2 + 2 x y + 2 y^2 - 84  x^2 y^2) +
         t(2 t^2 - 6t + 1) ) \\
         -\, 4 (3t-1)(32t-27)  y^3 ( y + 2 x  )
         +\,   (2t-3)(27t-16)  x^3 ( x  + 4  y )
          \end{array} \right) }}
\\
  && \times \Big({2880\,t\,(3t-1)(2t-3)(2 t^2-1)}\Big)^{-1}
\;.
\end{array}
\]
The parametrisation of $c,h$ and $w$ is invariant under
\be
  (t,x,y) \mapsto (\frac{1}{2t}, -\frac{y}{t}, -\frac x{2t} )
\;.
\label{eq:bswap}
\ee
As a result, whenever $t = p/p'$ is rational, we can, and henceforth
always will, take $p'$ odd.

\subsection{Structure of $\WB$ vacuum representations and the quotient
$\cF$} 
\label{sec:wba}

We are primarily interested in the space of fields spanned by normal
ordered combinations of $L$, $W$ and their derivatives which, as in
the case of the $\WA$ algebra, are in one-to-one correspondence with
the states in the quotient of the $h=w=0$ Verma module by the space
generated by $L_{-1}\vac$ and $W_{-1}\vac$ (here $L_{-1}\vac=
W_{-1}\vac=0$ forces $W_{-2}\vac=W_{-3}\vac=0$). We shall again call
this quotient space  $\cF$. Using the method of \cite{vddv}, we now
systematically search for all singular vectors in $\cF$. In the
subsequent section, we will check whether they do in fact give
conserved densities for some integrable perturbations.  

For all $t$, the vacuum representation corresponds to the weight
$\lambda_{vac}$: 
\[
  \lambda_{vac} + \rho
= (4t -2 )\omega_0 + (1-2t)\omega_1 + (1-t)\omega_1 = [4t-2,1-2t,1-t]
\;.
\]
The first step is again to find the simple roots of the set
$\Delta_{\lambda_{vac},+}^{re}$,  to identify  $W_{\lambda_{vac}}$ and
then to find the dominant weight in the shifted $W_{\lambda_{vac}}$
orbit of $\lambda_{vac}$. Again, it is straightforward to see that we
will only get singular vectors in $\cF$ for $t$ rational. When $t$ is
rational and positive, $W_{\lambda_{vac}} \equiv W$, the full Weyl
group of $c_2^{(1)}$, and  the dominant weight $\lambda$ in the orbit
$W\cdot\lambda_{vac}$ can be put in the form 
\[
  \lpp
= w ( \Lambda^+ - t \Lambda^- )
\;.
\]
$w$ is an element of the finite Weyl group generated by $w_{\alpha_1}$
and $w_{\alpha_2}$, $\Lambda^+$ is a dominant integral weight of
$c_2^{(1)}$ at level $p$, and $\Lambda^-$ is a dominant integral
weight of $d_3^{(2)}$ (the dual of $c_2^{(1)}$)  at level $p'-1$. This
means that  
\[
  \Lambda^+ = [\Lambda_0^+, \Lambda_1^+,\Lambda_2^+],
\;\;\; \sum_{i=0}^2 \Lambda_i^+ = p,\;\;\;\;\;\;
  \Lambda^- = [\Lambda_0^-, 2\Lambda_1^-,\Lambda_2^-],
\;\;\; \sum_{i=0}^2 \Lambda_i^- = p'-1,
\;\;\;
\]
and the  $\Lambda^\pm_i$'s are non-negative integers. We denote a
representation with such a weight $\lambda$ by 
$[ \Lambda^+_1,\Lambda^+_2 ; \Lambda^-_1,\Lambda^-_2 ]$. From the
determinant formula \cite{KWat2}, we can see that there are singular
vectors in the Verma module of  
$[\Lambda^+_1,\Lambda^+_2;\Lambda^-_1,\Lambda^-_2 ]$ at levels 
$\Lambda_1^+\Lambda_1^-$, $\Lambda_2^+\Lambda_2^-$ and
$\Lambda_0^+(\Lambda_0^- + 1)$.
Then, exactly as for the $\WA$ algebra, following the general
conjecture of \cite{vddv}, the composition series for all the Verma
modules with highest weights in $W \cdot \lambda_{vac}$ are given in
terms of the Kazhdan-Lusztig polynomials for the double coset 
\[
  W_{{\bar\Lambda}^-} \backslash W / W_{\Lambda^+}
\]
where
$ W_{\Lambda^+}$ is generated by the $\hs i$ for which
$\Lambda^+_i$ are zero, and $W_{{\bar\Lambda}^-}$ is generated by the
$\hs i$ for which $\Lambda^-_i, i=1,2$ are zero.

The outcome of these calculations, which are straightforward, is as
follows: the vacuum representation of $\WB$ with $t$ rational and
positive, $p'$ odd, falls into one of nine different classes presented
in table \ref{tab:wbvac}. There are singular vectors in $\cF$ only
in cases \rn 8 and {\rn 9}, which will now be considered in more
detail.

\begin{table}

\label{tab:wbvac}
{
\renewcommand{\arraystretch}{1.5}
\[
\begin{array}{|l| l l | l l | l l|}
\hline
   & p & p' & \Lambda^- & \Lambda^+ &  W_{{\bar\Lambda}^-} &
W_{\Lambda^+} \\
\hline
\rn 1  & 1 & 1 & [ 0\; 0\; 0]         & [ 0\; 1\; 0 ]
               & \{\hs 1, \hs 2\} & \{\hs 0, \hs 2\} \\

\rn 2  & 2 & 1 & [ 0\; 0 \;0]         & [ 1\; 1\; 0 ]
               & \{\hs 1, \hs 2\} & \{ \hs 2\} \\

\rn 3  & >2& 1 &  [ 0\; 0\; 0]        & [ 1\; 1\; p\!-\!2 ]
               & \{\hs 1, \hs 2\} &  \\

\rn 4  & 1 & 3 & [0\; 1\; 0 ]         & [0\; 1\; 0 ]
               & \{\hs 2\}        & \{\hs 0, \hs 2\} \\

\rn 5  & 2 & 3 & [0\; 1\; 0 ]         & [1\; 1\; 0 ]
               & \{\hs 2\}        & \{\hs 2\} \\

\rn 6  & >2& 3 & [0\; 1\; 0 ]         & [1\; 1\; p\!-\!2 ]
               & \{\hs 2\}        &  \\

\rn 7  & 2 &>3 & [p'\!-\!4\; 1\; 1 ]      & [0\; 1\; 1 ]
               &                  &  \{\hs 0 \}\\

\rn 8  & >2&>3 & [p'\!-\!4\; 1\; 1 ]      & [p\!-\!2\; 1\; 1 ]
               &                  &  \\

\rn 9  & 1 &>3 & [p'\!-\!4\; 1\; 1 ]      & [0\; 1\; 0 ]
               &                  & \{\hs 0, \hs 2\} \\
\hline
\end{array}
\]
\caption{The classification of the $\WB$ algebra vacua for $t$
rational and positive}
}
\end{table}

\subsubsection{\rn 8: Minimal models}
\label{sec:wcmin}

Models of type \rn 8, for which  $p \geq 3$, $p' \geq 5$, $p'$ odd,
correspond to the minimal models of the $\WB$ algebra. These are
conjectured to have representations parametrised by four integers,
$[ab;cd]$, with
\[
  x = a - 2t c
\;,\;\;\;
  y = b - t d
\;,\;\;\;\;
 0 < a,b,c,d ;\;\;\; (a+b) < p \;;\;\;\; (2c + d) < p'
\;.
\]
and to have a single highest-weight vector in $\cF$, at
level $(p-2)(p'-3)$, of type $[2p-3,1;1,1]$.

\subsubsection{{\rn 9}: A non-minimal model}

\begin{figure}
\label{fig:ix}
{
\[
\setlength{\unitlength}{0.01in}%
\begingroup\makeatletter\ifx\SetFigFont\undefined
\def\x#1#2#3#4#5#6#7\relax{\def\x{#1#2#3#4#5#6}}%
\expandafter\x\fmtname xxxxxx\relax \def\y{splain}%
\ifx\x\y   
\gdef\SetFigFont#1#2#3{%
  \ifnum #1<17\tiny\else \ifnum #1<20\small\else
  \ifnum #1<24\normalsize\else \ifnum #1<29\large\else
  \ifnum #1<34\Large\else \ifnum #1<41\LARGE\else
     \huge\fi\fi\fi\fi\fi\fi
  \csname #3\endcsname}%
\else
\gdef\SetFigFont#1#2#3{\begingroup
  \count@#1\relax \ifnum 25<\count@\count@25\fi
  \def\x{\endgroup\@setsize\SetFigFont{#2pt}}%
  \expandafter\x
    \csname \romannumeral\the\count@ pt\expandafter\endcsname
    \csname @\romannumeral\the\count@ pt\endcsname
  \csname #3\endcsname}%
\fi
\fi\endgroup
\begin{picture}(190,265)(35,535)
\thinlines
\put(132,780){\line( 0,-1){ 35}}

\put(125,730){\line(-2,-3){ 20}}
\put(140,730){\line( 2,-3){ 20}}

\put( 90,685){\line(-3,-4){ 25}}
\put(100,685){\line( 5,-6){ 25}}
\put(160,685){\line(-5,-6){ 25}}
\put(170,685){\line( 3,-4){ 25}}

\put( 65,635){\line( 0,-1){ 35}}
\put( 65,635){\line( 3,-2){ 50}}
\put(140,635){\line( 3,-2){ 50}}
\put(195,635){\line( 0,-1){ 35}}
\put(185,635){\line(-3,-2){ 50}}
\put(130,635){\line( 0,-1){ 35}}
\put(125,635){\line(-3,-2){ 50}}

\put( 65,580){\line(-1,-1){ 25}}
\put( 65,580){\line( 1,-1){ 25}}
\put(195,580){\line(-1,-1){ 25}}
\put(195,580){\line( 1,-1){ 30}}
\put(130,580){\line( 5,-2){ 80}}
\put(130,580){\line( 1,-1){ 25}}
\put(130,580){\line(-1,-1){ 25}}
\put(130,580){\line(-5,-2){ 80}}

\put(128,785){\makebox(0,0)[lb]{\smash{\SetFigFont{12}{14.4}{rm}1}}}
\put(128,730){\makebox(0,0)[lb]{\smash{\SetFigFont{12}{14.4}{rm}2}}}
\put(90,690){\makebox(0,0)[lb]{\smash{\SetFigFont{12}{14.4}{rm}3}}}
\put(163,690){\makebox(0,0)[lb]{\smash{\SetFigFont{12}{14.4}{rm}4}}}
\put( 60,640){\makebox(0,0)[lb]{\smash{\SetFigFont{12}{14.4}{rm}5}}}
\put(128,640){\makebox(0,0)[lb]{\smash{\SetFigFont{12}{14.4}{rm}6}}}
\put(192,640){\makebox(0,0)[lb]{\smash{\SetFigFont{12}{14.4}{rm}7}}}
\put( 60,585){\makebox(0,0)[lb]{\smash{\SetFigFont{12}{14.4}{rm}8}}}
\put(128,585){\makebox(0,0)[lb]{\smash{\SetFigFont{12}{14.4}{rm}9}}}
\put(190,585){\makebox(0,0)[lb]{\smash{\SetFigFont{12}{14.4}{rm}10}}}
\put( 30,535){\makebox(0,0)[lb]{\smash{\SetFigFont{12}{14.4}{rm}*}}}
\put( 95,535){\makebox(0,0)[lb]{\smash{\SetFigFont{12}{14.4}{rm}*}}}
\put(160,535){\makebox(0,0)[lb]{\smash{\SetFigFont{12}{14.4}{rm}11}}}
\put(225,535){\makebox(0,0)[lb]{\smash{\SetFigFont{12}{14.4}{rm}*}}}
\end{picture}
\]

}
\caption{Embedding pattern for the type \rn 9 module}
\end{figure}
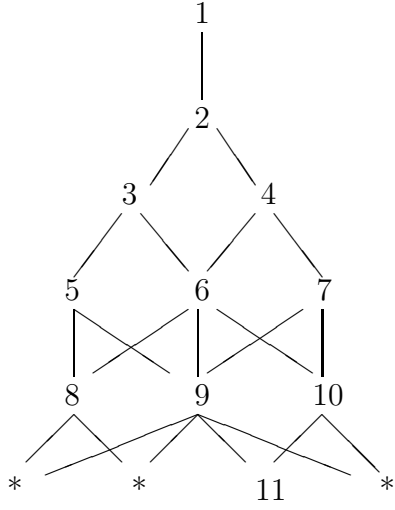

In this case, $\lambda_{vac}$ is not dominant.  The dominant weight 
$\lambda$ in $W \cdot \lambda_{vac}$ gives a representation with 
$h=2 - 2p'$.  From table \ref{tab:wbvac}, we see that the coset of
interest is $W/\{\hs 0,\hs 2\}$. The affine Weyl group $W$ has the
three generators $\hs 0,\hs 1,\hs 2$, with relations 
\be
  \hs i^2 = e         \;,\;\;
  \hs 0 \hs 2 = \hs 2 \hs 0 \;,\;\;
  \hs 0 \hs 1 \hs 0 \hs 1 = \hs 1 \hs 0 \hs 1 \hs 0 \;,\;\;
  \hs 2 \hs 1 \hs 2 \hs 1 = \hs 1 \hs 2 \hs 1 \hs 2 \;,\;\;
\;.
\label{eq:wrels}
\ee
The words in the Weyl group of length up to 5 are (with $i\equiv
\hs i$):
\def\q{,\,}
\[
\begin{array}{c}
        e \\
0 \q    1 \q    2 \\
01 \q 02 \q 10 \q 12 \q 21 \\
010 \q 012 \q 021 \q 101 \q 102 \q 121 \q 210 \q 212 \\
0102 \q 0121 \q 0210 \q 0212 \q 1010 \q 1012 \q 1021 \q 1210 \q 1212
\q 2101 \q 2102 \\
01021 \q 01210 \q 01212 \q 02101 \q 02102 \q 10102 \q 10121 \q
10210 \q 10212 \q 12101 \q 12102 \q 21012 \q 21021 \\
\end{array}
\]
We can now give the coset representatives $\underline{y}$ of minimal
length less than or equal to 5 of the coset $W/\{\hs 0,\hs 2\}$
\def\q{,\;}
\[
\begin{array}{c}
        e \\
        1  \\
01     \q 21 \\
101    \q 021   \q 121  \\
2101   \q 1021  \q 0121  \\
12101  \q 21021 \q 01021 \q 10121 \\
\end{array}
\;.
\]
All these elements must terminate with a 1, but we can further rule
out those words ending with a 1 that can be related through the
relations (\ref{eq:wrels}) to another element not ending with a 1: for
instance, $[2121] \sim [1212] \sim [121]$.
These are embedded as shown in Fig. \ref{fig:ix}.

We are especially interested in the Verma module of the vacuum and in
table \ref{tab:polys} give the minimal $\underline y$ and maximal
$\bar y$ coset representatives and the Kazhdan-Lusztig polynomials  
$P_{\bar x,\bar y}(q)$ for all submodules of $M_{\lambda_{vac}} =
M_{01 \cdot \lambda}$, which can be easily calculated using the
recursion formulae in 
\cite{vddv}. 
\begin{table}
\[
{\small
\begin{array}{c|ccccccccccc}
label  & 1    & 2     & 3     & 4      & 5      & 6    & 7
       & 8    & 9     & 10    & 11  \\
\hline
h      &2-p'  & 3-p'  & 0     & 4-p'   & 1      & 1    & 5-p'
       & 3    & 4     & p'-1  & 2p'-2   \\
\underline y
       & e    & 1     & 01    & 21     & 101    & 021  & 121
       & 2101 & 1021  & 0121  & 01021\\
\bar y & 02    & 102     & \! 0102 \!    & 2102     & 10102   & 02102   & 12102
       & 210102 & 102102 & 012102  & 0102102 \\
\hline
3   & & & 1 & & 1 & 1 & & 1 & 1 & 1 & 1 + q \\
5   & & & 0 & & 1 & 0 & & 1 & 1 & 0 & 1 \\
6   & & & 0 & & 0 & 1 & & 1 & 1 & 1 & 1 \\
8   & & & 0 & & 0 & 0 & & 1 & 0 & 0 & 0 \\
9   & & & 0 & & 0 & 0 & & 0 & 1 & 0 & 1 \\
10  & & & 0 & & 0 & 0 & & 0 & 0 & 1 & 1 \\
11  & & & 0 & & 0 & 0 & & 0 & 0 & 0 & 1 \\
\end{array}
}\]
\caption{Coset representatives and Kazhdan-Lusztig polynomials for
case \rn{9}}
\label{tab:polys}
\end{table}
The number of times $L(i)$ appears in the composition series of $M(j)$
is given by the value of the corresponding polynomial at $q=1$,
calculated for the maximal coset representative.  Repeating the
arguments of section \ref{ssec:lm}, we find that there is a singular
vector corresponding to node (11) in $\cF$, which has $h = 2p'-2$
and which transforms in the $[12;11]$ representation.

\subsection{Searching for integrable perturbations}
\label{sec:seven}

We now consider the perturbations of the minimal models in the
$\WB$ extended conformal theories.  As in the $\WA$ case,
we first consider which fields may occur in the ope
\be
 \PHI efgh \;\;\PHI abcd  \sim \sum \PHI{a'}{b'}{c'}{d'}
\;.
\label{eq:wope3}
\ee
where $\PHI efgh $ is the conserved density and $\PHI abcd$ the
perturbation. We first ensure that the coefficient of $\PHI abcd$ on
the r.h.s. of (\ref{eq:wope3}) is zero, which will restrict our
choices of $[ab;cd]$. Secondly, since we look for a singular conserved
density, that is, with $\PHI efgh$ a polynomial in $L$ and $W$, we may
restrict the sum in the r.h.s. of (\ref{eq:wope3}) to descendants of
$\PHI abcd$. We first consider the minimal models. 

\subsubsection{Integrable perturbations for type \rn 8 models}

The candidate conserved current is 
$\PHI{2p-3,}{1}{1}{1} \equiv \PHI 11{p'-2}1$.
To ensure that the coefficient of $\PHI abcd$ on the r.h.s of
(\ref{eq:wope3}) is zero we take the perturbing field $\PHI abcd$ to
be one of the minimal model fields, and consequently the only non-zero
singular descendant of the perturbing field is 
$\PHI{2p-a-2b,}bcd \equiv \PHI ab{p'-c-d,}d$.  We find  
\be
  \Delta h
= (p-a-b)(2c+d-3) + (p'-2c-d)(a+b-2) + (2c+d-3)(a+b-2)
\geq 0
\;.
\ee
Again, imposing $\Delta h = 2$, we find the solutions:
\be
\begin{array}{r|cccccc}
    & p & p'& a & b & c & d \\ \hline
(a) &   & 4 & 3 & 1 & 1 & 1 \\
(b) &   & 4 & 2 & 2 & 1 & 1 \\
(c) &   & 4 & 1 & 3 & 1 & 1 \\
(d) &   & 5 & 2 & 1 & 1 & 1 \\
(e) &   & 5 & 1 & 2 & 1 & 1 \\
(f) & 3 &   & 1 & 1 & 1 & 3 \\
(g) & 3 &   & 1 & 1 & 2 & 1 \\
(h) & 4 &   & 1 & 1 & 1 & 2
\end{array}
\ee
If no entry is given for $p$ or $p'$, that variable is free.
We see that in cases $(a)$, $(b)$ and $(c)$ that $p'$ is even, which
contradicts our assumptions.  Therefore, we only need to check whether
the coefficient $\mu$ in (\ref{eq:wope}) vanishes for cases
$(d),(e),(f),(g)$ and $(h)$. 

Let us look at the singular descendant of the perturbing field
which occurs in the r.h.s. of the o.p.e. (\ref{eq:wope3}), :
{\renewcommand{\arraystretch}{1.5}
\be
\begin{array}{l|ccccc}
    &  (d) & (e) & (f) & (g) & (h) \\ \hline
\PHI{a}{b}{c}{d}
&
\PHI{2}{1}{1}{1} &
\PHI{1}{2}{1}{1} &
\PHI{1}{1}{1}{3} &
\PHI{1}{1}{2}{1} &
\PHI{1}{1}{1}{2}  \\

\PHI{2p-a-2b,}bcd
&
 & 
 & 
 & 
\PHI{3}{1}{2}{1} &
\PHI{5}{1}{1}{2} \\

\PHI ab{p'-c-d,}{d}
&
\PHI{2}{1}{3}{1} &
\PHI{1}{2}{3}{1} &
\PHI{3}{1}{1}{3}
 & 
 & 
\end{array}
\;.
\ee}%
Of this  list, $(d)$ and $(g)$ are guaranteed to have a total
derivative at the first order pole, as they have a level--1 singular
vector in the Verma module of the field on the r.h.s. of the ope. The
argument is the same as in the $\WA$ case: since the Verma module has
only two states at level 1, namely 
\be
  L_{-1} \vec \psi
\;\hbox{ and }\;
  W_{-1} \vec \psi
\;,
\ee
whenever there is a singular vector at level 1, $W_{-1}\vec\psi$  can
be expressed in terms of $L_{-1}\vec\psi$, so that any state at level
1 can be written as a total derivative. The only possible exception is
when $L_{-1}\vec\psi$ is itself the singular vector, but this occurs
only for the vacuum representation, and in that case there are no
states at all at level one. 

In the other three cases, $(e),(f)$ and $(h)$, we can repeat the
analysis of appendix \ref{sec:wacoeffs}, and find that these do 
not, in fact, correspond to integrable perturbations.

We now examine $(d)$ and $(g)$ in turn:

\subsubsection{$(d):\;\;[21;11]$ perturbation }

For the $[21;11]$ perturbation, there is a conserved current
transforming in the $[11;31]$ representation when $p'=5$, in which
case the conserved current has weight $(2p-4)$. $p$ must be coprime to
$5$, which gives the values of the spins $\Delta$ of the conserved
charges as $(2p-5)$, 
\[
\begin{array}{c|cccccccc}
p       & 3 & 4 & 6 & 7 & 8 & 9 &  \cdots\\ \hline
\Delta  & 1 & 3 & 7 & 9 &11 &13 &  \cdots
\end{array}
\]
This covers the complete set of the  $a_4^{(2)}$ exponents.

\subsubsection{$(g):\;\;[11;21]$ perturbation}

For the $[11;21]$ perturbation, the conserved current transforms
in the $[31;11]$ representation when $p=3$, in which case the
conserved current has weight $(p'-3)$. $p'$ must be odd and coprime to
$3$, i.e., $p'$ coprime to 6: This gives the values of the spins
$\Delta$ of the conserved charges as
\[
\begin{array}{c|ccccccc}
p'     & 5 & 7 & 11 & 13 & 17 & 19 & \cdots\\ \hline
\Delta & 1 & 3 &  7 &  9 & 13 & 15 & \cdots
\end{array}
\]
This is not the complete set of exponents  of any affine  extension of
$c_2$. It is clear that this perturbation should correspond to either
the $c_2^{(1)}$ or  $d_3^{(2)} $  Toda theory. Hence, either some of
the conserved currents are never null, or we simply fail to see when
they are null by this method. 

\subsubsection{Integrable perturbations for type {\rn 9} models}

For type {\rn 9} models ($p=1$), the candidate conserved current is
$\PHI 1211$, of weight $2p'-2$. Since these are not minimal models, we
would have to identify those  perturbing fields that may be
considered, but we  shall restrict ourselves to  perturbing fields of
the form $\PHI 11ab$. The singular descendants of the conserved
current at levels 1 and 2 ensure  \cite{bajnok} that the only possible
operator products are   
\[
  \PHI 1211 \;\; \PHI 11ab \;\sim\; \PHI 12ab
\;,
\]
which gives
\[
  \Delta h
= a + b - 2 = 2
\qquad \Rightarrow\qquad
  (a,b) = (3,1), (2,2), (1,3)
\;.
\]
For the third possibility, the presence of a level-1 singular vector
descendant of $\PHI 1213$ ensures the conservation of $\int \PHI
1211$. The other two cases are ruled out as before. 

For this $[11;13]$ perturbation, the conserved current transforms
in the $[12;11]$ representation when $p=1$, in which case the
conserved current has weight $(2p'-2)$, where $p'=(2k+1)\geq 5$.  The
values of the spins $\Delta$ of the conserved charges are then
\[
\begin{array}{c|cccccccc}
p'     & 5 & 7 &  9 & 11 & 13 & 15 & \cdots\\ \hline
\Delta & 7 & 11& 15 & 19 & 23 & 27 & \cdots
\end{array}
\]
This is not the complete set of exponents of any affine extension of
$c_2$. It is clear that this perturbation should correspond to either
the $c_2^{(1)}$ or  $d_3^{(2)} $  Toda theory. Hence, as before,
either some of the conserved currents are never null, or they cannot
be probed  by this method.

\subsection{Discussion}

We have examined the $\WB$ space $\cF$ of fields which are normal
ordered polynomials of $L$, $W$ and their derivatives and have
identified the cases for which such a field may be a null field. These
fall into two classes, namely the $(1,p')$ models and the $(p,p')$
models for $p' \geq 5$. In the first case, we have found that the
$[11;13]$ representation is integrable with conserved quantities of
weight $2p'-2$ for $p'$ odd, transforming in the $[12;11]$
representation, and leading to a set of conserved charges whose
weights lie in subset of the exponents of $c_2^{(1)}$. In the second
case, we have found that the $[21;11]$ perturbation is integrable for
$p'=5$ with conserved densities of weight $2p-4$ transforming in the
$[11;31]$ representation, leading to conserved quantities whose
weights are all the exponents of $a_4^{(2)}$.  In addition, we found
that the $[11;21]$ perturbation is integrable for $p=3$ with conserved
densities of weight $p'-3$ transforming in the $[31;11]$
representation, leading to conserved quantities with weights which are
a subset of the exponents of $c_2^{(1)}$.   Thus our analysis exhausts
the perturbations which are known to be integrable from the results of
\cite{FF}: we find the perturbations  $\{[11;21],[31;11],[21;11]\}$ to
be integrable for particular values of $c$, and accordingly 
their duals,  $\{[12;11],[11;13],[11;12]\}$ are also integrable. 

\section{Explicit calculations of conserved charges}
\label{sec:nine}

We have examined several possible perturbations using the computer.
For each perturbation we have examined whether there are conserved
quantities small values of $\Delta$. Whenever we found a conserved
density, we then examined in which cases it was a singular vector in
$\cF$. The results of these investigations are summarised in the
tables below. 

In the case of each perturbation we have two equivalent
parametrisations of the highest weights, related by $t \to 1/t$ for
the $\WA$ algebra and $t \to 1/(2t)$ for the $\WB$ algebra, and in
each case we choose that parametrisation for which the 
perturbation is of type $[11;rs]$ for integer labels $r,s$.
Consequently, the values of $t$ for which we find the perturbation has
a singular conserved density will not always agree with the values of
$t$ in the body of the paper above, but the identifications are
straightforward.
As every perturbed conformal theory for which we found conserved
quantities can be thought of as an affine Toda theory, we have
labelled the perturbation by the Toda theories to which it
corresponds.   

The entries of the tables as as follows:

An `X' indicates that there is no conserved quantity of the given 
spin $\Delta$ for the given perturbation for {\bf generic} values of
$c$. From the calculations we have performed it is not possible to
tell if there is a finite set of values of $c$ for which there is a
conserved quantity. 

A `$-$' indicates that there is a conserved quantity of the given
spin $\Delta$ for the given perturbation for {\bf generic} values of
$c$, but that there is no value of the central charge for which the
conserved current is a highest-weight vector. 

In the remaining cases, we have found that there is a conserved
quantity of the given spin $\Delta$ for the given perturbation for
{\bf generic} values of $c$, and that for one or more values of the
central charge, the  conserved current is a highest-weight vector in
$\cF$. In these cases we give $t$, $c$, and the eigenvalues $h$
and $w$ of the {\bf perturbing} field.

\subsection{Perturbations of $\WA$ models}

In each case we parametrise the weights of the perturbing field as
$h_{[11;rs]}(t), w_{[11;rs]}(t)$ for integer $r$, $s$. These are
identical to 
$h_{[sr;11]}(1/t), w_{[sr;11]}(1/t)$

{
\renewcommand{\arraystretch}{1.5}
\[
\begin{array}{||c||c|c|c||}
\hline \hline
\Delta
& \begin{array}{l}
a_2^{(1)} \\
h_{[11;22]} = 3 t - 2 \\
w_{[11;22]} = 0
\end{array}
&
\begin{array}{l}
g_2^{(1)},
d_4^{(3)} \\
h_{[11;12]} = \frac{4}{3}t - 1 \\
w_{[11;12]} =
  \frac{ (3 - 4t)(3 - 5t) }{ 27 t^{1/2} }
\end{array}
&
\begin{array}{l}
g_2^{(1)},
d_4^{(3)} \\
h_{[11;14]} = 6 t - 3 \\
w_{[11;14]} =

\frac{ (1 - 2t)(2 - 3t) }{ t^{1/2} }
\end{array}
\\[1mm] \hline\hline
  2
&
  \left\{
  {\renewcommand{\arraystretch}{1.1}\begin{array}{l}
   t = 5/3 \\ c=-22/5 \\ h = 3 \\ w = 0 \\
   \end{array}} \right.
  \left\{
  {\renewcommand{\arraystretch}{1.1}\begin{array}{l}
   t = 3/5 \\ c = -22/5 \\ h = -1/5 \\ w = 0 \\ 
   \end{array}} \right.
&
  {\rm X}
&
  {\rm X}
\\ \hline
  3
&
 {\rm X}
&
  {\rm X}
&
  {\rm X}
\\ \hline
  4
&
  \left\{
  {\renewcommand{\arraystretch}{1.1}\begin{array}{l}
   t = 7/3 \\ c = -114/7 \\ h = 5 \\ w = 0 \\
   \end{array}} \right.
  \left\{
  {\renewcommand{\arraystretch}{1.1}\begin{array}{l}
   t = 3/7 \\ c = -114/7 \\ h = -5/7 \\ w = 0 \\
   \end{array}} \right.
&
  {\rm X}
&
  {\rm X}
\\ \hline
  5
&
  \left\{
  {\renewcommand{\arraystretch}{1.1}\begin{array}{l}
   t = 8/3 \\  c = -23 \\h = 6 \\ w = 0 \\
   \end{array}} \right.
  \left\{
  {\renewcommand{\arraystretch}{1.1}\begin{array}{l}
   t = 3/8 \\  c = -23 \\h = -7/8 \\ w = 0 \\ 
   \end{array}} \right.
&
  \left\{
  {\renewcommand{\arraystretch}{1.1}\begin{array}{l}
   t = 2/3 \\ c=-2 \\h = -1/9 \\ w = - \sqrt 6 / 486 \\
   \end{array}} \right.
&
  \left\{
  {\renewcommand{\arraystretch}{1.1}\begin{array}{l}
   t = 3/2 \\ c=-2 \\ h = 6 \\ w = - 7 \sqrt{ \frac 23}
   \end{array}} \right.
\\ \hline
  6
&
  {\rm X}
&
  {\rm X}
&
  {\rm X}
\\ \hline
  7
&
  \left\{
  {\renewcommand{\arraystretch}{1.1}\begin{array}{l}
   t = 10/3 \\ c=-186/5 \\ h = 8 \\ w = 0 \\ 
   \end{array}} \right.
  \left\{
  {\renewcommand{\arraystretch}{1.1}\begin{array}{l}
   t = 3/10 \\ c=-186/5 \\ h = -11/10 \\ w = 0 \\ 
   \end{array}} \right.
&
  -
&
  -
\\ \hline
\end{array}
\]
}

\subsection{Perturbations of $\WB$ models}

In each case we parametrise the weights of the perturbing field as
$h_{[11;rs]}(t), w_{[11;rs]}(t)$ for integer $r$, $s$. These are
identical to 
$h_{[sr;11]}(1/(2t)), w_{[sr;11]}(1/(2t))$.  
Consequently, although in section \ref{sec:seven} we considered the
integrable perturbations of type $[31;11]$ and $[12;11]$, we have not
presented independent data for these, as it can be read of from that
given for the $[31;11]$ and $[11;21]$ perturbations.

{
\renewcommand{\arraystretch}{1.4}
\[
\begin{array}{||c||c|c||}
\hline
\Delta
& \begin{array}{l}
c_2^{(1)} , d_{3}^{(2)} \\
h_{[11;13]} = 6t - 3 \\
w_{[11;13]} =
- \frac {\left(6 t - 5\right)\left(3 t - 1\right)
         \left(2 t - 1\right )\left(7 t - 3\right)}
        {30 \left(2 t - 3\right)\left(2 t^2 -1 \right)}
\end{array}
&
\begin{array}{l}
c_2^{(1)} , d_{3}^{(2)} \\
h_{[11;21]} = 4 t - 2 \\
w_{[11;21]} =
  \frac{ (2t-1)(7t - 3)(8t - 5) }
       { 90 ( 2 t^2 - 1) }
\end{array}
\\[1mm] \hline\hline
  2
&
  {\rm X}
&
  {\rm X}
\\ \hline
  3
&
  \left. {\renewcommand{\arraystretch}{1.1}
  \begin{array}{l}
   t = 7/6  \\ c = -68/7 \\ h = 4 \\ w = 1 \\ 
  \end{array}}\right.
&
  \left. {\renewcommand{\arraystretch}{1.1}
  \begin{array}{l}
  t = 3/7  \\ c = -68/7 \\ h = -2/7 \\ w = 0 \\
  \end{array}} \right.
\\ \hline
  4
&
  {\rm X}
&
  {\rm X}
\\ \hline
  5
&
  -
&
  -
\\ \hline
  6
&
  {\rm X}
&
  {\rm X}
\\ \hline
  7
&
  \left\{
  {\renewcommand{\arraystretch}{1.1}
   \begin{array}{l}
   t = 11/6 \\ c = -444/11 \\ h = 8 \\ w = - 3186/515 \\ 
   \end{array}} \right.
  \;,
  \left\{
  {\renewcommand{\arraystretch}{1.1}
   \begin{array}{l}
   t = 1/5 \\ c = -76 \\ h = -9/5 \\ w = -152/7475  \\ 
   \end{array}} \right.
&
  \left\{
  {\renewcommand{\arraystretch}{1.1}
   \begin{array}{l}
   t = 3/11 \\c = -444/11 \\ h = -10/11 \\ w = 62/3399 \\ 
   \end{array}} \right.
\;,
  \left\{
  {\renewcommand{\arraystretch}{1.1}
   \begin{array}{l}
   t = 5/2  \\ c = -76 \\ h = 8 \\ w = 58/69 \\ 
   \end{array}} \right.
\\ \hline\hline
\Delta
&
\begin{array}{l}
a_4^{(2)} \\
h_{[11;12]} = \frac{5}{2}t  - \frac 32 \\
w_{[11;12]} =
-\frac {\left(5 t - 3\right)\left(8 t - 5\right)
        \left(5 t - 4\right)\left(7 t - 3\right )}
       {720\, \left(2 t - 3\right)\left(2 t^2 - 1 \right)}
\end{array}
&
\begin{array}{l}
{a_4^{(2)}} \\
h_{[11;31]} = 10 t - 4 \\
w_{[11;31]} =
  \frac{ (22t - 5)(5t - 3)(5t - 2)}
       { 90 (2 t^2 - 1) }
\end{array}
\\[1mm] \hline\hline
  2
&
  {\rm X}
&
  {\rm X}
\\ \hline
  3
&
  \left.
  {\renewcommand{\arraystretch}{1.1}
   \begin{array}{l}
   t = 5/8 \\  c = 1/2 \\ h = 1/16 \\ w = 0 \\ 
  \end{array}} \right.
&
  \left.
  {\renewcommand{\arraystretch}{1.1}
   \begin{array}{l}
   t = 4/5 \\  c = 1/2 \\ h = 4 \\ w = 1 \\ 
  \end{array}} \right.
\\ \hline
  4
&
  {\rm X}
&
  {\rm X}
\\ \hline
  5
&
  {\rm X}
&
  {\rm X}
\\ \hline
  6
&
  {\rm X}
&
  {\rm X}
\\ \hline
  7
&
  \left.
  {\renewcommand{\arraystretch}{1.1}
   \begin{array}{l}
   t = 5/12 \\ c = -11 \\ h = -11/24 \\ w = -253/1055808 \\ 
  \end{array}} \right.

&

  \left.
  {\renewcommand{\arraystretch}{1.1}
   \begin{array}{l}
   t = 6/5 \\ c = -11 \\ h = 8 \\ w = 214/141 \\ 
  \end{array}} \right.

\\ \hline
\end{array}
\]
}

\subsection{Discussion}

The first observation we can make from these results is that
they are in complete agreement with those predicted by the
singular-vector analysis.  We find singular conserved densities at
exactly the weights and central charges predicted.

Secondly,  in some cases, although there is a conserved current for
all values of $c$, it is in fact never a highest-weight vector for
either choice of the perturbing field of a dual pair. This is unlike
the Virasoro perturbations, for which there is always some value of
the central charge for which one field of a dual pair is a
highest-weight vector in the vacuum module.

The third observation is that the conserved quantities always come in
pairs.  If there is a perturbation with a conserved current of weight
$\Delta$ at central charge $c$, then there is another perturbation at
the same central charge, also with a conserved current of weight
$\Delta$, for which the perturbing field itself has weight
$\Delta+1$. On inspection, the conserved current is the same in both
cases, and the perturbing field in the second case is simply the
conserved current itself: if $J$ is a highest-weight vector in the
vacuum, then we clearly have $ [ \oint J , \oint J ] = 0$ and so $J$
is a conserved current ensuring the integrability of the (irrelevant)
perturbation by the field $J$. 

Another interesting observation is that for the $[11;13]$ and $[11;21]$
perturbations of the $\WB$ models, there are two values of the
central charge for which the conserved density of spin 7 becomes
singular. This is an unusual feature,  encountered here for the first
time.

\section{Conclusions}

The singular vector analysis presented here provides a remarkably
simple probe of the integrability of perturbed conformal field
theories.  The first step essentially boils down to identifying dual
pairs of fields in the minimal models under consideration.  This
already tells us that, in the theory perturbed by one of the pair,
something nontrivial  is conserved (namely the other member
of the pair). The second step is a little refinement: by identifying
the conserved current with a descendant of the identity, we
demonstrate that what is conserved is actually local.  All this is
obtained without any explicit computation of the conserved
charges. When supplemented by duality, and by allowing the conserved
charges to be either singular or subsingular, this argument yields all
the known integrable perturbations, at least for the three class of
models analysed here. It would be interesting to test this method
further and in particular to apply it to superconformal theories.  We
will report on this in a separate work.

Have we learned something on the corresponding quantum integrable
systems?  As already indicated, our analysis gives a strong
integrability signal.  But since these models are nothing but quantum
affine Toda theories, for which integrability has already been proven
in \cite{FF}, what is the point? At first, we should stress that the
integrability  proof of Feigin and Frenkel  is extremely abstract.
Our argument provides a much simpler and intuitive way of seeing
integrability.  But more importantly,  the argument of \cite{FF}  does
not provide any constructive devise for obtaining the conservation
laws.  The singular vector analysis gives us explicit expressions for
the conserved charges, even though each of them correspond to a
different value of $c$. But for this to be a genuine handle on a
constructive understanding of integrability, we would have to answer
the truly interesting question: How to find a one-parameter
deformation of these densities that preserves their conservation?

\section{Acknowledgements}

GMTW was supported by an NSERC (Canada) International
Fellowship and an EPSRC (UK) advanced fellowship.

The work of PM was supported by NSERC (Canada).

We are grateful to P. Di Francesco for his collaboration in the early stages
of this work. GMTW would like to thank K. de Vos and P. van Driel for many 
very helpful discussions of their work.

All computer calculations were performed in MAPLE using the cft Maple
Package written by H.G.~Kausch \cite{maple}, and some were performed
on computers purchased on EPSRC (UK) grant GR/J73322.

\vspace{1cm}

\setcounter{section}{0}
\def\thesection{\Alph{section}}

\section{Free-field representations of $W$-algebras, integrable
perturbations and affine Toda theories}
\label{sec:six}
\label{app:toda}

In this appendix, we recall the relation of affine Toda theories to
integrable perturbations of the general class of $W\bar g$ algebras
which can be represented by $r$ free bosons and characterised by a
dual pair of  finite dimensional Lie algebras 
${\bar g},\,{\bar g}^\vee$ of rank $r$. This of course includes the
Virasoro case, for which ${\bar g}={\bar g}^\vee=a_1$, but also the
$\WA$ and $\WB$ algebras to be analysed later.

If we denote the $r$ free bosons by the $r$ -component vector $X(z)$,
then the $W$ algebras' free-field representatives are the commutants
of the screening charges:  
\[
\int dz~ \exp( i {\sqrt t} \alpha_j \cdot X )
\;\;,\;\;\;\;
\int dz~ \exp( - \frac{i}{{\sqrt t}|\alpha_j|^2} 2 \alpha_j \cdot X )
\;\;,\;\;
j = 1 ,2
\]
where $\{\alpha_j\}$ is the set of the simple roots of the Lie
algebras $\bar g$ and $\{\alpha_j^\vee=2\alpha_j /|\alpha_j|^2\}$ is
the set of simple roots of its dual $\bar g^\vee$.  Each of the two
sets of screening charges can also be thought of as parts of the
Hamiltonian of a Toda theory in the lightcone framework (the first set
corresponds to the conformal Toda theory based on the algebra 
${\bar g}$, and the second set to the conformal Toda theory based on
the dual algebra ${\bar g}^\vee$.) The standard normal-ordered
exponential vertex operators 
\be
: \exp(i\bm\cdot X(z) ) :
\;
\label{eq:ffpf}
\ee
transform as primary fields of the $W\bar g$-algebras.

Formally, the affine Toda theory based on an affine algebra $g$ is the
perturbation of the $W\bar g$ conformal Toda theory by the term
\be
  \int d^2z \; \exp\left(\, 
  i \sqrt t \alpha_0\cdot( X(z) + \bar X(\bar z) ) \,\right)
\;,
\label{eq:p1}
\ee
where $\alpha_0$ is the extra root of the algebra $g$.
As a consequence, this perturbation can be identified with a
perturbation by a primary field of the $W\bar g$ algebra.
The conserved densities for this perturbation are equally well
conserved densities for $g^\vee $ affine Toda theory, viewed as the
perturbation of the $W\bar g^\vee $ model by the term
\be
  \int d^2z \; \exp\left(\, 
  - 2 i \frac{\sqrt t}{\alpha_0^2} 
   \alpha_0\cdot( X(z) + \bar X(\bar z) ) \,\right)
\;.
\label{eq:p2}
\ee
As a result, each perturbation of a $W\bar g$ can be related to up to
two affine Toda theories. Furthermore, since the perturbing fields 
(\ref{eq:p1}) and (\ref{eq:p2}) are dual to each other, they each
provide non-local conserved densities for the affine Toda theory with
the other perturbation. We have examined the cases when these
non-local conserved densities can be identified with local polynomials
in the W algebra fields.

The W algebras found by Hamiltonian reduction are exactly those which
have a free-field representation. The free-field representative of the
$W\bar g$ primary field obtained by Hamiltonian reduction of the $g$
highest-weight representation $\lambda$ is of the form (\ref{eq:ffpf})
with 
\be
  \bar\mu 
= \frac{1}{\sqrt t} 
  \left( \bar\lambda + t \bar\rho^\vee   \right)
\;,
\label{eq:bm}
\ee
and thus we can identify the integrable perturbations we have
considered in this paper with affine Toda field theories.

\subsection{Free-field representations and dual pairs for $\WA$}
\label{ssec:a2}

The free-field representation \cite{FZ} provides a description of the
primary fields but only for the  minimal models, for which $t$ is a
rational number: $t=p/p'$, with $p$, $p'$ coprime and greater than 3.
This restriction to minimal models is adequate for our present purpose
which is to identify dual pairs.  As before, a minimal model primary
field can be  parametrised by a set of 4 strictly positive integers
$[ab;cd]$ subject to $a+b< p$ and $c+d< p'$, with 
\[
  \lambda+\rho = [p-a-b,a,b] - t[p'-c-d,c,d]
\;.
\]
{}From (\ref{eq:bm}), the finite weight $\bm $ appearing in the vertex
representation $e^{i\bm\cdot X}$ of the field $\phi_{[ab;cd]}$ is 
\[
\bm = \left( (a-1) t^{-1/2} - (c-1) t^{1/2} \right) {\bar \omega}_1
    + \left( (b-1) t^{-1/2} - (d-1) t^{1/2} \right) {\bar \omega}_2
\;,\;\;
\]
In an orthogonal basis, the $ a_2$ fundamental weights ${\bar \omega}_i$ are
\[
{\bar \omega}_1 = (\frac 1{\sqrt{2}}, \frac 1{\sqrt 6})
\;,\;\;
{\bar \omega}_2 = (0,\sqrt{\frac 23})
\;.
\]
The complete set of pairs $\{\bm,2\bm/|\bm|^2\}$ with both members
labelled in terms of strictly positive integers $[ab;cd]$ for all
values of $t$ are: 

\[
\{  [ 22;11] , [11;22] \}
\;,\;\;
\]\[
\{  [12;11]  , [11;14] \}
\;,\;\;
\{  [21;11]  , [11;41] \}
\;,\;\;
\{  [11;12]  , [14;11] \}
\;,\;\;
\{  [11;21]  , [41;11] \}
\;.
\]

The first case corresponds to the $a_2^{(1)}$ self-dual theory, and
the next four to the $\{g_2^{(1)},d_4^{(3)}\}$ dual pairs of affine
Toda theories. These are the dual pairs we were looking at. The Toda
field theory argument shows that each member of such a pair is an
integrable perturbation.  In our approach, we want to establish the
integrability by a  singular-vector argument. We stress again that  by
taking one of the two fields as the perturbation, the other is a
conserved density of the corresponding massive theory,  albeit not
necessarily local.  

\subsection{Free-field representations  and dual pairs for $\WB$}

The free-field construction for the $\WB$ minimal models is given in
\cite{KWat2}.  Primary fields are parametrised by 4 strictly positive
integers $[ab;cd]$ satisfying $a+b<p,\;2c+d<p'$.  The finite weight of
the corresponding vertex operator is 
\[
\bm = \left( (a-1) t^{-1/2} - 2(c-1) t^{1/2} \right) {\bar \omega}_1
    + \left( (b-1) t^{-1/2} -  (d-1) t^{1/2} \right) {\bar \omega}_2
\;,\;\;
\]
where $t=p/p'$. In an orthogonal basis, the $c_2$ fundamental weights
${\bar \omega}_i$ are 
\[
{\bar \omega}_1 = (\frac 1{2}, \frac 1{2})
\;,\;\;
{\bar \omega}_2 = (0,1)
\;.
\]
The complete set of integrable dual pairs, in  the $[ab;cd]$ notation is:
\[
\{  [12;11]  , [11;13] \}
\;,\;\;
\{  [31;11]  , [11;21] \}
\;,\;\;
\{  [13;11]  , [11;12] \}
\;,\;\;
\{  [21;11]  , [11;31] \}
\;.
\]
The first two correspond to the $\{ c_2^{(1)}, d_3^{(2)} \}$ dual
pairs and the third and fourth to the $a_4^{(2)}$ self-dual affine
Toda theory.

\section{Calculating the coefficients in the ope
of two $\WA$ primary fields.}

\label{app:wope}
\label{sec:wacoeffs}

Let us  consider three $\WA$ primary fields, $\Phi_a$, $\Phi_b$,
$\Phi_c$, (with corresponding states $\vec a$, $\vec b$, $\vec c$) and
the state $\vec\Psi=\left(\lambda L_{-1}+\mu W_{-1}\right)\vec a $.
Taking various inner products, we find 
\be
  \pmatrix{ \cev a L_1 \vec\Psi \cr \cev a W_1 \vec\Psi }
= \pmatrix{   2 h_a &  3 w_a \cr
              3 w_a & \frac{1}{48} h_a( 32 h_a + 2 - c) }
  \pmatrix{\lambda \cr \mu}
\;,
\label{eq:zero}
\ee
so that
\be
  \pmatrix{\lambda \cr \mu}
= \frac1{\rm det}
  \pmatrix{ \frac{1}{48} h_a( 32 h_a + 2 - c) & - 3 w_a \cr
            - 3 w_a                           &   2 h_a }
  \pmatrix{ \cev a L_1 \vec\Psi \cr \cev a W_1 \vec\Psi }
\;,
\label{eq:one}
\ee
where
\be
  {\rm det}
= \frac{1}{24} h_a^2( 32 \, h_a + 2 - c) - 9 w_a^2
\;.
\ee
{}From the o.p.e.s of a $\WA$ primary field,
\begin{eqnarray*}
  L(z) \vec\Phi
& = & \left(\, z^{-2} h \vec\Phi + z^{-1} L_{-1} \right) \vec\Phi + O(1)
\\
  W(z) \vec\Phi
& = & \left(\, z^{-3} q \vec\Phi + z^{-2} W_{-1}
                             + z^{-1} W_{-2} \right) \vec\Phi + O(1)
\end{eqnarray*}
we get the commutation relations:
\be
  [L_m - L_{m-1} , \Phi(1) ] = h \, \Phi(1)
\;,\;\;
  [W_m - 2 W_{m-1} + W_{m-2} , \Phi(1) ] = w \, \Phi(1)
\;.
\ee
With these relations, we obtain
\be
\pmatrix{
  \cev a L_1 \Phi_b \vec c \cr
  \cev a W_1 \Phi_b \vec c }
 =
\pmatrix{
  ( h_a + h_b - h_c )   & 0 \cr
  ( 2 w_a + w_b - 2 w_c) & 1 }
\pmatrix{ \cev a \Phi_b \vec c \cr \cev a \Phi_b W_{-1} \vec c }
\;.
\label{eq:two}
\ee
To find $\cev a \Phi_b W_{-1} \vec c $, we can use the null vectors in the
representation $c$. There are two cases of interest:

\begin{itemize}

\item{}
Cases (a) and (b): $\Phi_c = \PHI 1n11$

In this case there is a null vector in representation $c$ at level 1:
\be
  \left(\,W_{-1} - \frac{3 w_c}{2 h_c} L_{-1} \,\right) \vec c
\;,
\ee
so that
\be
  \cev a \Phi_b W_{-1} \vec c
= \frac{3 w_c}{2 h_c} ( h_b + h_c - h_a )
  \cev a \Phi_b \vec c
\;.
\ee

\item{}
Case (c): $\Phi_c = \PHI 2211$.

In this case there is a null vector at level 2,
\be
  \left(\, W_{-2} - \frac{2}{h_c + 1} L_{-1} W_{-1} \,\right)\vec c
\;,
\ee
and using the commutation relations before, we find
\be
  \cev a \Phi_b W_{-1} \vec c
= -\frac{ (h_c + 1) }{2}
   \frac{( w_a - w_b)}{( h_a - h_b )}
  \cev a \Phi_b \vec c
\;,
\ee
remembering that in this case $w_c = 0$. As a result, we find that
\be
  \mu =
\cases{
  \frac{ (n-1) p^{1/2} p'{}^{3/2}}
       {(np' + 3p - 2pp')(np' + 2p - p' - pp')}
  & Cases (a) and (b) \cr
0 & Case (c) }
\ee

\end{itemize}

\section{Zhu's algebra for the non-minimal $\WA$ model}
\label{app:zhu}

The way in which null vectors in the vacuum representation of a chiral
algebra restrict the field content is by now well understood, from
Feigin and Fuchs' work for the Virasoro algebra \cite{FeFu3},
Zhu's work in the general case \cite{zhu}, and \cite{mecmp} for the
particular case of the $\WA$ algebra.

The `minimal models' of a certain chiral algebra are characterised by
the presence of a `sufficient' number of singular vectors in $\cF$
such that only a finite number of representations are consistent.
However, the $p'=2$ models of the $\WA$ algebra are not believed to be
minimal:  a continuous set of representations of the $\WA$ algebra are
consistent with the vanishing of the null fields in $\cF$. According
to \cite{mecmp}, we should be able to find all the constraints on the
allowed representations $\{h,w\}$ by considering the constraint that
the three-point functions 
\be
  \langle h,w | \; \Phi_{h,w} \; W_{-1}^a \psi \;=\; 0
\;,
\label{eq:tpf}
\ee
vanish, where $\psi$ runs over the highest-weight vectors in $\cF$.
These three-point functions are polynomials in $h$ and $w$.
Whether this is a full set of constraints on $\{h,w\}$ or not, they
are certainly necessary.

We have investigated these constraints for the $(3,2)$ and $(5,2)$
models, being the simplest of the $p'=2$ models. In these, we have
found two singular vectors in $\cF$, both at level $3p-3$, i.e. levels
$6$ and $12$ respectively. The constraints arising inserting these two
vectors, and their $W_{-1}$ descendents, are found to be identical.
They are: 
\be
\begin{array}{c | c }
p & \hbox{constraints} \\
\hline
3 & 54 w^2 - 8 h^3 - h^2 = 0 \\[2mm]
5 & \begin{array}{c}
    (270 w^2 - 40 h^3 - 27 h^2) \\
    ( 1350 w^2 - 200 h^3 - 495 h^2 - 408 h - 112) = 0 \\
    \end{array} \\
\end{array}
\ee
Looking for solutions of the form $h=h[c,d,1,1], ~w=w[c,d,1,1]$,
these constraints become
\be
\begin{array}{c | c c }
p & \hbox{constraints} \\
\hline
3 & \begin{array}{c}
    ~~(c-1)(c-2) (d-1)(d-2) \\
    \times\;
    (2c+2d-5)(2c+2d-7) = 0\\
    \end{array}\\[5mm]
5 & \begin{array}{c}
    ~~(c-1)(c-2)(c-3)(c-4) (d-1)(d-2)(d-3)(d-4) \\
    \times\;
    (2c+2d-7)(2c+2d-9)(2c+2d-11)(2c+2d-13) =0  \\
    \end{array} \\
\end{array}
\ee
It is a simple conjecture that the allowed values of $\{h,w\}$ in the
$(p,2)$ models are $\{h=h[c,d,1,1], w=w[c,d,1,1]\}$, where
one of the following holds:
$ c = 1,2 \ldots p-1$,
$ d = 1,2 \ldots p-1$,
$ 2(c+d) = p+2,p+4 \ldots 3p-2$.
This justifies the above choice for the form of $h$ and $w$.

\end{document}